\documentclass{article}
\usepackage{graphicx} % Required for inserting images
\usepackage{xcolor} % Required for inserting images
\usepackage{subfig}
\usepackage{natbib}
\usepackage{makecell}
\usepackage{appendix}
\usepackage{authblk}
\usepackage[a4paper, total={6in, 8in}]{geometry}
\providecommand{\keyword}[1]
{
  \small	
  \textbf{\textit{Keywords --- }}% #1
}
\title{When tiny convective spread affects a midlatitude jet: spread sequence}
\author[1,2]{Edward Groot}
\author[2]{Michael Riemer}
\affil[1]{Currently: Atmospheric, Oceanic and Planetary Physics, University of Oxford, UK. Email: large.edward.simulations@gmail.com}
\affil[2]{Institut für Physik der Atmosphäre, Johannes Gutenberg University, Mainz, Germany. Email: mriemer@uni-mainz.de}
\date{February 2025}

\begin{document}

\maketitle
\begin{abstract}
    We investigate the evolution of spread over three days in a numerical ensemble experiment starting from tiny initial condition uncertainty.
    %with an ICON-ensemble (with rescaled initial uncertainty of ECMWF's IFS, forecasting 3 days ahead). 
    We simulate %The experiment considers a real event, in which 
    a real event during which three mesoscale convective systems occur in close proximity to the midlatitude jet. 
    Combining ensemble sensitivity analysis with a spread-growth diagnostic based on piecewise potential-vorticity tendencies,
    we compare the spread evolution with an existing conceptual three-stage model.
    %it can be shown that each of these convective systems imposes different effects on the jet stream.
    %
    %The spread-growth pattern is compared with a three-stage conceptual model \citep{Zhang2007,Baumgart2019}. The latter study suggested that (according to statistically averaged diagnostics), latent heating variability in convective systems perturb divergent winds in their outflows. The variability in these winds may constructively interact with PV-gradients, followed by a barotropic (Lorenz-type) spread production along the jet stream, which amplifies large-scale balanced jet stream perturbations within a similar idealised ensemble.
    %
    Each system follows the first stage, characterised by development of convective variability. Nevertheless, we find significant variation among the %three mesoscale convective
    systems in their propensity to interact with 
    the jet stream, which characterises the second conceptual stage. 
    %a strong PV gradient. 
    One exemplary convective system follows the conceptual evolution of \citeauthor{Baumgart2019}, i.e., convective uncertainty initially projects onto the jet by upper-tropospheric outflow, which further amplifies spread through balanced nonlinear %(\citeauthor{Lorenz1969}-type) 
    growth as it propagates downstream. 
    %For one convective system spread evolves following the conceptual sequence, i.e., initially driven by small convective heating perturbations (equivalent to about $\approx$ 0.1 mm area mean precipitation accumulation). One day later, it 
    Rossby-like dispersion in the downstream spread is strongly associated with the convective variability. % system and downstream shift of the jet stream is demonstrated. 
    In contrast, for another convective system, convective variability projects onto the local anticyclonic flow aloft only. %, i.e., by introducing potential vorticity spread in the rotational component of the anticyclonic wind.
    Subsequently, this anticyclonic perturbation does not project considerable convective uncertainty onto the particularly straight jet stream, which truncates the conceptual spread evolution. %, spread in which
    %The second conceptual stage of spread growth is thereby associated with the adjustment to balance of the convective outflow, as hypothesized by \citeauthor{Zhang2007}. 
    For the third system, negligible fingerprints of distinct spread growth stages (beyond the initial stage) are identified.

    Alongside convective heating, 
    %W
    %we find that 
    longwave radiation 
    jointly %plays a leading-order role for 
    dominates the spread evolution near the convective systems, whereas earlier studies suggest convective heating dominates. %(\citeauthor{Baumgart2019,Selzetal2021}).
    Longwave-radiative tendencies of convective anvils outlive the accompanied heating tendencies and extend spatially.
    Furthermore, we link convective variability of the exemplary system directly to longwave-radiative tendencies.
    Therefore, longwave radiation appears to contributes substantially to stages 1 and 2 here.

    %On the side, the spread-growth tendencies also reveal that longwave-radiative heating near the tropopause may play an important role in the first (convective-scale) spread-growth stage.
    %
    %The two other convective systems indicate spread development associated with the first stage of the conceptual sequence of spread growth, but the second and third stage never dominate. Signals of the onset of stage 2 (and in at least one case also stage 3) are rather weak and remain local. 
    Finally, we identify flow dependence of the impact of convection on the jet, which may relate to the wave-relative location of convective systems. % relative to a wavy jet.
    %We speculate that their location with respect to a ridge, located over continental Europe, and their distance from the gradient may play a role here. 
    Hence, we advocate to improve understanding of particularly favourable conditions for %upscale and 
    downstream propagation of convective variability. %remains an important topic for further investigation.
    %, and in particular the conditions that tend to support this upscale and downstream spread propagation.    
    %Lastly, our analysis suggests that tiny perturbations in mesoscale average precipitation accumulation (on the order 0.1 mm) can trigger Rossby-like dispersion about a day later.  
\end{abstract}
\newpage
\section{Introduction}
Atmospheric predictability is ultimately limited by the butterfly effect \citep{Lorenz1969,Palmer_2014}, i.e., the rapid growth of forecast spread at small scale and their impact on successively increasing scales. Current-day operational forecasts have not yet reached this limit \citep[e.g.,][]{judt2018, selz_2019, zhangetal2019, Selzetal2021} and a better understanding of the “butterfly effect” helps to assess the remaining improvement potential. 
%Despite substantial improvements over the last decades \citep{bauer2015}, current-day operational forecasts have not yet reached this limit . 
%To assess the remaining improvement potential, it is of importance to better understanding the upscale growth of forecast errors that underlies the proverbial ''butterfly effect''. 
The seminal work by \cite{Lorenz1969} describes this upscale spread growth in terms of a two-dimensional model, in which scale interactions are described as if the atmosphere were fully developed two-dimensional turbulence (with continuous upscale and downscale interaction). This model thereby assumes self-similarity across scales, implying that the underlying spread-growth mechanism is independent of scale. More recent work, however, identified that moist convection, which is unrepresented in Lorenz’ model, is of crucial importance for rapid spread growth at small scales \citep[e.g.][]{hohenegger2007,Zhang2007,Selz2015a} .

Moving beyond Lorenz’ assumption of self-similarity of spread growth, \cite{Zhang2007} introduced a three-stage conceptual model with each stage being dominated by distinct growth mechanisms. \cite{Baumgart2019} analysed spread-growth mechanisms quantitatively and confirmed a three-stage sequence of upscale spread growth, but with some differences in the interpretation of the dominant spread-growth mechanisms. Baumgart et al.’s quantification of spread-growth mechanisms is based on the potential-vorticity (PV) perspective of forecast spread \citep{davies2013diagnosis}. From this PV perspective, the pattern of forecast errors, or forecast spread, in the midlatitudes is dominated by the PV errors in the tropopause region. A distinct advantage of a PV perspective is that standard piecewise PV thinking for midlatitude dynamics \citep[e.g.][]{davis1996,TR21} can be applied to identify individual spread-growth mechanisms \citep{Baumgart2018,Baumgart2019,binder2021,BR19,Selzetal2021}.
%, which generalizes to the spread and errors of forecasts, i.e., ensemble spread \citep{BR19,Selzetal2021}. 
In addition, the approach has proven useful to yield insight into differences in the dynamical evolution that arise from differences in the representation of physical processes \citep[specifically demonstrated for cloud-radiative impacts on idealized cyclones;][]{behrooz}. 

The three-stage sequence of spread growth in \cite{Baumgart2019} has been identified in the mean behaviour of a number of spread-growth experiments (distributed evenly over a year) when averaged over large domains. From this mean perspective, spread growth is initially (in the first 12 h) dominated 
by the stochastic (deep) convection scheme, in the subsequent 36 h by upper-tropospheric divergent flow, and finally by the rotational (non-divergent) flow. This picture holds for initial condition uncertainty with very small amplitude. Increasing initial condition uncertainty, a transition in the relative importance of spread-growth mechanisms is observed \citep{Selzetal2021}. For current-day initial condition uncertainty, spread growth at initial time is on average already dominated by the rotational wind. This transition in the relative importance of spread-growth mechanisms provides further evidence for the idea that the intrinsic limit of predictability is indeed characterised by the three-stage sequence of spread growth \citep{Selzetal2021}.
The dominance of the convection scheme during the first stage is consistent with earlier notions of the importance of latent heat release in convection for rapid small-scale spread growth when convection is resolved \citep{hohenegger2007,Zhang2007,Selz2015a}. The progression to divergent winds as mediator during the second stage is consistent with the idea that a general adjustment-to-balance-process projects spread from precipitation regions to larger scales: differences in latent heat release leads to differences in the divergent component of the wind and in the subsequent (geostrophically) balanced state on the inertial time scale of approx.\ 12h \citep {Zhang2007,Selz2015a,Bierdel2017,Bierdel2018}. \cite{Baumgart2019}, however, hypothesise a more effective principal pathway for the divergent winds to project spread upscale: that differences in the divergent, convective outflow leads to differences in displacements of the sharp PV gradient associated with the midlatitude jet, and thus to large-amplitude, meso-scale PV spread. In the third stage, these spread further grow by quasi-barotropic nonlinear interaction \citep{Baumgart2018,Baumgart2019}, consistent with the spread-growth mechanism in the \cite{Lorenz1969} model. 

The mean perspective on spread growth provided by \cite{Baumgart2019} and \cite{Selzetal2021} suggests an intriguing picture: Variability in latent heat release in convective systems generates variability in the upper-tropospheric divergent winds, which directly project variability onto the large PV gradient associated with the jet stream. 
%Accordingly, 
%outflow of convection 
The sensitivity of upper-tropospheric PV above convective systems and the subsequent evolution of the downstream flow to the representation of convective systems is well known \citep[e.g.,][]{doneetal2006, clarkeetal19, clarkeetal2019, Rodwell2013, lojkoetal2022}. %; this is particularly true for  convection-permitting versus convection-parameterised configurations \citep[e.g.][]{, ,,Grootetal2023}. 
Furthermore, %purely from the PV-perspective, 
mesoscale convection is known to create PV dipole patterns with negative anomalies downshear and above heating anomalies \citep[e.g.][]{chagnon2009,weijenborg2015,Weijenborgetal2017,Shutts2017}, providing further theoretical ground to study interactions between convective-scale flow and the large-scale flow from the PV perspective. 
However, a dominant role of upper-tropospheric divergent outflow impinging on the jet has not (yet) been demonstrated explicitly.
Such a dominant role is well established for other weather systems with strong latent heat release, tropical cyclones or warm conveyor belts, that interact with the jet \citep[e.g.][]{riemer_etal_2008,grams2011,keller2019}. 
For convective systems, it is well known that convective variability, in particular in terms of convective organization and amount of latent heat release, translates to variability in the divergent outflow aloft.
In numerical models, when grid spacing is coarse and therefore convection needs to be parameterised, outflow variability tends to depend linearly on variability in latent heat release \citep{Grootetal2023}.
When convection is represented more realistically, and arguably in the real atmosphere, outflow variability depends nonlinearly on latent heat release \citep{BrethertonSmolarkiewicz1989,Nichollsetal1991,Mapes1993,Groot_Tost_2022b,Grootetal2023}.
%Furthermore, recent work demonstrated that the variability in the magnitude of the deep convective outflow in ICON NWP simulations of the Munich hail storm event predominantly depends on linearly on latent heating rate, while grid spacing, convective aggregation and, possibly, further aspects of convective organisation explain the remaining variability \citep{Grootetal2023}. However, in large-eddy simulations (and presumably in reality), latent heating rate, convective aggregation and convective organisation is thought to be explain outflow variability \citep{Groot_Tost_2022b,BrethertonSmolarkiewicz1989,Nichollsetal1991,Mapes1993}.
To our knowledge, only one study so far examines explicitly the role of divergent outflow for the downstream impact of a mesoscale convective system \citep{lojkoetal2022}, but this study focuses on negative PV structures that are embedded in the outflow layer rather than displacement of PV gradients.
%have utilised forecast errors (derived from reanalysis) following an MCS-event in four NWP models. They reveal critical impact on a Rossby wave packet, causing downstream heat wave conditions \citep[similar to][]{Rodwell2013}. 

In the current study, we investigate the mechanisms which transition the initial uncertainties in convective systems into spread at large scales% and the balanced flow
thoroughly.
Thereby, it turned out to be impractical to follow the approach of \citet{Baumgart2019} and \citet{Selzetal2021} to use spatial composite of spread tendencies to describe the different contributions of each system to spread.
Instead, we will focus on spread tendencies through a thoughtful examination of the spatial patterns of spread tendencies.
Main questions that will be addressed are: Does the multi-stage sequence of mechanisms (which holds in the mean) also apply to individual convective systems? And thus, by extension: Is the advection of the PV gradient by the divergent flow indeed the most effective mechanism to project forecast spread onto the jet?

We address these questions by an ensemble spread-growth experiment covering a time period that encompasses the previously studied ”Munich Hail Storm” event \citep[e.g.][]{Wilhelmetal,Grootetal2023}. This period covers three simulated mesoscale convective systems that occur near the large PV gradient of the midlatitude jet over Europe in summer 2019. The ensemble is generated using initial condition uncertainty from the operational Integrated Forecast System (IFS). As in \cite{Selzetal2021}, the IFS' initial perturbations are re-scaled by a factor 1/1000. Accordingly, moist-convective processes dominate early-stage spread growth \citep[similar to ][]{Selzetal2021}. Essentially, our setup ensures very fast initial divergence of the ensemble in regions of mesoscale convection and is thus suitable to analyse the upscale impacts arising from these convective regions. 

Our analysis utilizes, besides the PV diagnostic of \cite{Baumgart2019} and \cite{Selzetal2021}, ensemble sensitivity analysis (ESA), combining the method with ideas from \cite{Grootetal2023}. % and Groot and Tost 2023a along the lines of Bednarczyk and Ancel, 2015. > details to be added in methods!
For coarse grid spacing (here: 13 km) and parameterised (deep) convection, mesoscale convection’s outflow is nearly proportional to its precipitation rate and thus linearly correlated \citep{Grootetal2023}, which translates to a linear correlation of the associated variability. In such a setup, ESA can exploit the linear correlation to quantitatively link the ensemble’s outflow variability with that of the precipitation rate (as a proxy of latent heat release). To make use of this quantitative link, the design of our spread-growth experiment judiciously includes a deterministic scheme to parameterize deep convection. Deterministic convection schemes underestimate spread growth \citep{Selzetal2021,plantcraig2008} and we thus expect that our ensemble experiment is underdispersive. Despite this caveat, and for the purpose of our current study, we consider the use of a deterministic scheme to be helpful for the interpretation of 
%process interactions that underly 
the mechanisms governing (upscale) spread growth.

In Section 2, we will summarise the methods of ensemble sensitivity analysis and potential vorticity spread tendency diagnostics. In Section 3 we will introduce the meteorological setting of our experiments and the key convective systems. Furthermore, we will introduce the spread evolution at near-tropopause levels for this event. 
%In Section 4, we extensively discuss the key insights that our analysis provides – the main results. 
Section 4 presents our results, followed by a discussion (Section 5) %of how we interpret the outcomes of this study 
and a final summary (Section 6).

\section{Methods}
\label{methods}
\subsection{Simulation setup}
\label{setup}
The selected episode is studied in the Icosahedral Nonhydrostatic (ICON) model at 26 km grid spacing over a global domain. A nest with 13 km grid spacing is used over Europe, with a two-way feedback between the nest and the global domain \citep{zangl2015icon,giorgetta2018icon,icontutorial2019}. The configuration is largely the same as the deterministic run of the corresponding setup in \cite{Grootetal2023} (see table \ref{tab:settings} for details), using mostly the standard parameterisation schemes (except for radiation, for which the Ritter-Geleyn scheme is used). The simulations here, however, have been initialised based on initial conditions of the IFS-ensemble on June 10$^{th}$ 2019, 12 UTC. We run an ensemble of 50 perturbed members with one additional unperturbed member.  All members are integrated out to 75~h lead time (~h denotes hours lead time, from here on). Simulation output is stored every 30 minutes. 
The initial condition uncertainty is taken from the corresponding 50-member IFS ensemble, but has been reduced by a factor 1000 (constant in space), following \cite{Selzetal2021}. The total ensemble spread is thus equal to 0.1\% of the original (operational) forecast uncertainty. Thereby, our uncertainty regime closely resembles the intrinsic limit (\cite{melhauser_zhang_2012,SunZhang2016}), in which the whole upscale cascade and spread amplification (e.g. \cite{Selzetal2021,DurranGingrich2014,Lorenz1969}) is fully passed through over time (starting from the effective resolution). Thereby, it is assured that convective-scale variability 
considerably affects the uncertainty of the jet evolution, at least early on, and that thus 
%is at some point (in the first two simulation days, \citep{Baumgart2019,Selzetal2021}) likely to affect the uncertainty in downstream evolution of the jet stream considerably, so that 
the potential impact  of the mesoscale convective variability for our selected cases can be meaningfully quantified.
\begin{table*}[htbp]
\caption{\label{tab:settings}Key simulation settings. }
\begin{tabular}{|c|cc|}
\hline
\textbf{Domain}                                     & \textbf{Global domain}                                                                 & \textbf{European Nest}                                                                   \\ \hline
\textbf{Model version  }                                     & \multicolumn{2}{c|}{ 2.6.0}    \\ \hline
\textbf{Grid spacing (km) }                                  & 26 (R03B06)                                                                            & 13 (R03B07)                                                                                  \\ \hline
\textbf{Time step (s) }                                      & 100                                                                                    & 50                                                         \\ \hline
\textbf{Domain top altitude (km)  }                                 & \multicolumn{2}{c|}{ 75}                                                                                                                     \\
Number of vertical levels (-)                       & \multicolumn{2}{c|}{90}                                                                                                                                                                        \\ \hline
\textbf{Deep convection parameterisation  }                  & \multicolumn{2}{c|}{\cite{tiedtke1989}, \cite{bechtoldetal2014}  }                                                                    \\ 
\makecell[c]{Time step deep convection (s) \\and subgrid orography} & 1200                                                                                    & 600                                                                                   \\ 
Time step gravity wave drag (s)                     &\multicolumn{2}{c|}{1200}                                                                                                                                                                                       \\ \hline
\textbf{Microphysics parameterisation           }            &\multicolumn{2}{c|}{ 1M \citep{seifert2008}          }                             \\ \hline
\textbf{Radiation parameterisation }                         &\multicolumn{2}{c|}{ Ritter-Geleyn}                                  \\ 
Time step radiation (s)                             & \multicolumn{2}{c|}{1800}                                                                                                                                                                     \\ 
Grid spacing radiation (km)                         & 52                                                                                     & 26                                                                    \\ \hline
\textbf{Rayleigh damping height (km)   }                     &\multicolumn{2}{c|}{ 22}                                              \\ \hline
\textbf{Initial conditions   }                     & \multicolumn{2}{c|}{IFS analysis      }                                                                                                                                                                                                                         \\ 
 Ensemble  perturbations             & \multicolumn{2}{c|}{IFS ensemble,  rescaled by 0.001, }   \\  & \multicolumn{2}{c|}{after \cite{Selzetal2021}     }
 \\ \hline
\end{tabular}
\end{table*}
   \newline
The simulation output is stored at a 0.25 degrees (about 28 km) output grid within the European nest; for the analysis of potential vorticity structures and their tendency diagnostics, re-gridding to a corresponding 0.75 degree resolution is applied, by means of bi-linear interpolation using CDO \citep{cdo}.
\subsection{Diagnostic tools}
The two diagnostics used in this work are ensemble sensitivity analysis \citep{ancell2006,torn2008} and PV spread tendency diagnostics \citep{BR19}. These diagnostics complement each other and the combination of spatial structures in statistical signals and in PV tendencies % \color{red}and an approximate assessment of causal interference \color{black} 
strengthen our confidence in the diagnosed interactions between convective variability and the PV field. 
Our analysis focuses on near-tropopause dynamics. Hence, the potential vorticity spread tendencies are temporally averaged over 3 hour intervals and vertically over an isentropic layer with a thickness of 7.5 K, centered around $\theta = 331.25$ K. %(near the tropopause; climatologically within the lower range of the mean altitude of the maximum jet stream gradient for June, \cite{rothlisberger2018}). 
The ensemble sensitivity is carried out at 250 hPa. 
Analysed PV-fields (originally 13 km grid spacing) have been re-gridded to 0.75 degrees grid spacing for the analysis (approximately the effective resolution). 

The combination of tools allows us to identify the multi-stage uncertainty evolution over the course of a few days %in a weather situation over Europe
, starting near the intrinsic limit of initial uncertainty. Consequently, the role of processes in typical forecasting scenarios is not diagnosed. However, the role of convective variability in the downstream propagation of uncertainty can be assessed thoroughly.  %\newline

\subsubsection{Ensemble sensitivity analysis}
Ensemble sensitivity analysis (ESA) is essentially a regression technique, which correlates patterns at $(x + dx, t + dt)$ with presumably correlated precursors at $(x,t)$. Here, $dt$ and $dx$ can cover the full 4D atmospheric evolution (and $dt$ would therefore not strictly have to be positive, while it should be when assessing the plausibility of causal relations). The lagged correlation is assessed by regressing across ensemble members. Insight in the variability of a pattern can be gained by inspecting the evolution of the pattern and by investigating its relation to plausible statistical precursors. % as correlated with the precursors, over a time span. 
The slope of the regression lines (relative to the ensemble variance) expresses the estimated strength of the correlation between a pattern at $x$ and another at $x + dx$. This slope is usually displayed in consecutive maps. %A correlation does not yet necessarily imply a causal link; 
By statistical testing one may be able to distinguish correlation patterns of interest from random correlation patterns. Furthermore, understanding of  spatio-temporal evolution of these patterns and their relation to other diagnostics and/or relation to pre-existing hypotheses allows one to learn about their significance. Furthermore, a coherent set of ensemble sensitivity analyses can improve the insight in the chain of processes relevant for a certain pattern of variability (e.g. \cite{Groot_Tost_2022}). Further examples of investigations of convective systems using ESA are, for instance, found in  \cite{Hanleyetal2013,Bednarczyk_Ancell_2015,Torn_Romine_2015}.\newline
Ensemble sensitivity analysis will be applied to the magnitude of instantaneous mass divergence (i.e. outflow rate) in the upper troposphere, averaged over a box volume that covers the upper tropospheric part of a targeted convective system. 
Three hour time windows are used for the diagnosis of mass divergence to filter out high frequency variability in divergence signals and account for a time lag between mass divergence and downstream impact.
Because the PV-gradient is typically not (directly) co-located with the convective systems, it may take several hours for advected PV and gravity waves to reach the PV-gradient.% The averaging is also applied to account better for the time lag in between and remove short-lived local (noisy) variability. %\newline
%
%\textcolor{red}{I suggest to add here a brief mention of ESA using rain rates(?) instead of divergence and that both give consistent results. I further suggest to then omit the appendix and references to "rain rate ESA" in the results section, if still included.}
%
\subsubsection{Quantitative PV perspective on spread growth}
%Quantitative PV framework for spread-growth mechanisms}
%\color{blue}to be revised; following Baumgart and Riemer (2019).\color{black}
\label{sec:PV-persepctive}
   % PV tendency diagnostics can be used to investigate the processes behind the amplification of potential vorticity variance within an ensemble: growth of potential enstrophy. 

We consider \citet{ertel1942} PV, $q$, on isentropic levels:
\begin{equation}
    q = \frac{\zeta + f}{\sigma} \,,
\end{equation}
where $\zeta$ is the component of relative vorticity perpendicular to an isentropic surface, $f$ the Coriolis parameter, and $\sigma = -g^{-1}(\partial p/\partial \theta)$ the isentropic layer density with gravity $g$, pressure $p$, and potential temperature $\theta$.
The local rate of change of PV is
\begin{equation}
 \frac{\partial q}{\partial t} = - \mathbf{v} \cdot \mathbf{\nabla}q + \mathcal{N} \,,
 \label{eq:PV_tendency}
\end{equation}
where $\mathbf{v}$ is the horizontal wind, $\mathbf{\nabla}$ is the horizontal gradient on an isentropic level, and the nonconservative term $\mathcal{N}$
\begin{equation}
 \mathcal{N} := - \dot\theta\, \frac{\partial q}{\partial \theta} + q \, \frac{\partial \dot\theta}{\partial \theta} + \frac{\nabla\times\dot{\mathbf{v}}}{\sigma} \,,
 \label{eq:noncon_PV_tendency}
\end{equation}
with diabatic heating rate $\dot\theta$, and nonconservative sources and sinks of momentum $\dot{\mathbf{v}}$.
As in \citet{Selzetal2021}, we decompose the horizontal wind in its non-divergent ($\mathbf{v_{rot}}$) and irrotational ($\mathbf{v_{div}}$) components 
\begin{equation}
    \mathbf{v} = \mathbf{v_{rot}} + \mathbf{v_{div}} \,.
\label{eq:helmholtz_decompostion}    
\end{equation}
Our final partition of the PV equation can thereby be written as
\begin{equation}
 \frac{\partial q}{\partial t} = - \mathbf{v_{rot}} \cdot \mathbf{\nabla}q - \mathbf{v_{div}} \cdot \mathbf{\nabla}q + \sum_i \mathcal{N}_i \,,
 \label{eq:PV_tendency_partition}
\end{equation}
where $\mathcal{N}_i$ denotes the $i^{th}$ nonconservative process, e.g., as represented by a parameterisation scheme in a numerical model. 
The basic idea \citep{davies2013diagnosis} is to analyse the dynamics of the \emph{difference} between two atmospheric states by the differences of the individual terms of the respective PV equation.
As in \citet{BR19}, we identify one state with the ensemble mean (denoted by an overbar) and the other state with an individual ensemble member.
Indicating the deviation from the ensemble mean by $\Delta$, considering (half of) the square of the PV deviation as a positive definite metric, and rearranging terms \citep[see][for details]{Baumgart2019, Selzetal2021} we obtain 
\begin{equation}
    \frac{1}{2}\frac{\partial}{\partial t}(\Delta q)^2 = ADV_{rot} + ADV_{div} + \sum_{i} \Delta q \Delta \mathcal{N}_i + \mathcal{R} \,,
    \label{eq:tendencies}
\end{equation}
where $ADV_{rot}$ and $ADV_{div}$ denote the terms that are associated with advection by the non-divergent and irrotational wind, respectively.
These terms are given by
\begin{equation}
    ADV_{rot} = \Delta q (-\Delta \mathbf{v}_{rot} \cdot \nabla \overline{q})
\end{equation}
and
\begin{equation}
    ADV_{div} = \Delta q (  -\Delta \mathbf{v}_{div} \cdot \nabla \overline{q} + \frac{1}{2} \Delta q \, \nabla \cdot \overline{\mathbf{v}_{div}}) \,.
\end{equation}
Note that we have eliminated in this formulation the terms that merely redistribute $(\Delta q)^2$, instead of change its magnitude \citep[see][]{Baumgart2018}.
In addition, we have introduced a residuum $\mathcal{R}$, which encapsulates discretisation and interpolation errors, and the nonconservative impact of numerical diffusion, which is not captured by model parameterisation \citep{Baumgart2018}.

Based on Eq.~\ref{eq:tendencies}, it is straightforward to derive a tendency equation for ensemble variance \citep{BR19}. More recently, this variance framework has also been used by \cite{schmidt2024}. %\citet{Schmidt_etal_2025}.
Furthermore, note that the spread growth formulation in \citet{BR19} and \citet{Selzetal2021}, which is based on the sum over permutations of the differences between ensemble members, can be shown to be equivalent to a variance formulation.  
%Further following \citet{Baumgart2019} and \citet{Selzetal2021}, we define growth rates associated with the individual terms by dividing the associated tendency by the domain-average of ensemble variance.

In summary, we arrive at a tendency equation that quantifies the individual contributions to the evolution of ensemble variance associated with PV advection by the non-divergent wind (hereafter referred to as rotational tendency), associated with PV advection by the irrotational wind (hereafter referred to as divergent tendency), and the nonconservative tendencies due to parameterisation schemes.
We will explicitly present tendencies from the deep convection and the longwave-radiation scheme. Tendencies from the gridscale precipitation and shortwave-radiation schemes have been found to be of subordinate importance and will not been shown. Previous work \citep{Baumgart2019} found that tendencies from other parameterisation schemes (turbulence, gravity wave drag) are negligible compared to convection and have thus not been considered in this study. 

The tendency diagnostics are computed over the 13 km European nest only. The Spherical Harmonics package \textit{Windspharm} \citep{windspharm} for Python has been utilised to split the wind into rotational and divergent wind components, following Equation \ref{eq:helmholtz_decompostion}.
%\color{red}Introduce in a paragraph quantitative measures that are diagnosed from PV-tendencies, plus some kind of equation to give insight in the budget\color{black}
%\newline
%\color{blue}In this work, the ESA and PV tendency diagnostics are utilised to investigate the potential role of convective systems for ridge shifts and amplification over Europe during one event, using the simplification of a deterministic convection scheme, combined with rescaled IFS initial condition perturbations to represent error dynamics, which implies carefully designated idealisation of errors on the smaller scales. \color{black}
%
%
%
\section{Synoptic overview and initial PV spread}%To be determined}%Convective systems}
\subsection{Synoptic setting}
\begin{figure}
    \centering
    \includegraphics[width=160mm]{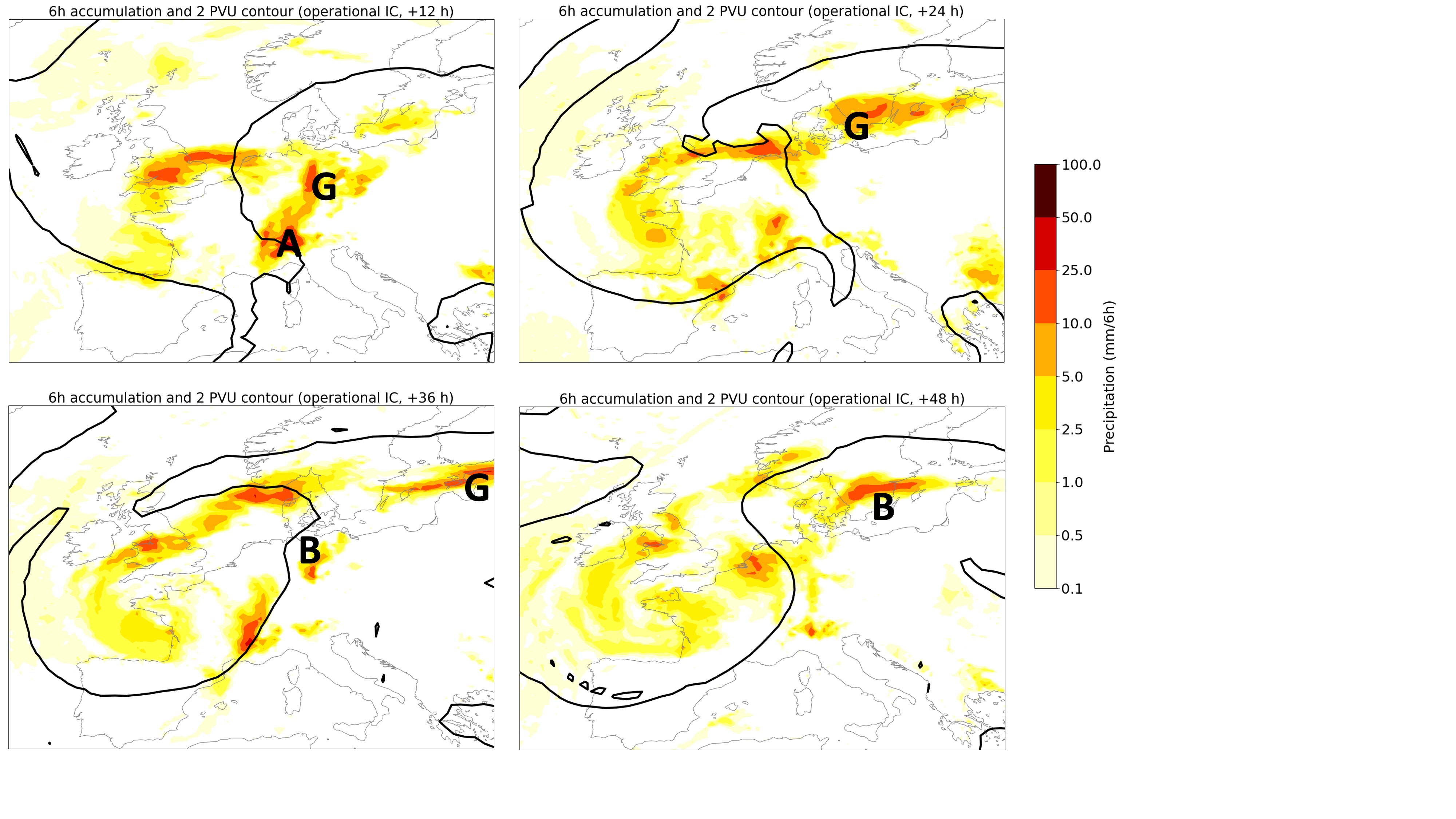}
    \caption{Accumulated precipitation over past 6~h for 12 (top left), 24 (top right), 36 (bottom left) and 48~h (bottom right) forecasts of ICON (using the ECMWF IFS operational run's initial conditions). The PV contour of 2PVU at 250 hPa is contoured in black (coarse grained to 0.75 degrees). Labels A, B and G mark the Alps convective system, the Baltic Sea convective system and the Germany convective system. }
    %\caption{Six hour ICON forecast of 700 hPa potential temperature along with mean sea level pressure isobars and the 250 hPa wind speed (represented as wind barbs) over Europe in the control simulation for June 10, 18 UTC. }
    \label{fig:synoptic-chart}
\end{figure}

Investigating the interaction of divergent convective outflow with the strong PV gradient associated with the midlatitude jet requires that mesoscale convective systems occur in proximity to that gradient. 
%The investigation of our hypothesis requires interaction between convective systems and the jet stream. 
From 10 -- 12 June 2019, a previously studied period \citep{Wilhelmetal, Grootetal2023}, convection presumably fulfills this requirement.
During this period, 
%Therefore, a previously investigated event in which mesoscale convection occurs over Europe is selected: June 10$^{th}$ to 12$^{th}$ 2019. In the upper troposphere, a large trough is located over France (PV contours in Fig. \ref{fig:synoptic-chart}a). Ahead of the trough, a 
southerly flow ahead of an upper-level trough over France advected warm and moist air over the European continent (not shown), %into Central Europe; this combination 
favouring the formation of deep convection. 
Three organized convective systems occurred ahead of the trough and within the downstream ridge, %in our ICON simulations, 
while the upstream trough underwent cyclonic wave breaking (Fig.~\ref{fig:synoptic-chart}).

On 10 June, one convective system occurred very close to the upper-level trough over the western Alps (hereafter referred to as Alps system; labeled "A" in Fig.~\ref{fig:synoptic-chart}a).
At the same time, a second system initiated farther downstream of the trough over Germany. This system subsequently moved northward into the Baltic Sea and then further downstream (hereafter referred to as the Germany system; labeled "G" in Fig.~\ref{fig:synoptic-chart}a,b,c). 
%The first deep convection in this region subsequently initiates over the slopes of the Western parts of the Alps on June 10th in the afternoon, as well as further northeast over the central parts of Germany (Fig. \ref{fig:synoptic-chart}a). 
%These convective systems, which have their main outflow at upper tropospheric levels \citep[about 200-300 hPa; see also ][]{Grootetal2023,grootthesis}, move northward over Germany during the night. 
%An air mass with intermittent convective instability is highly suitable for deep moist convection and as the accompanying baroclinic zone slowly advances eastward onto the continent, 
A third  mesoscale convective system initiated during the night of 11--12 June over Northern Germany, and subsequently moved northward over the Baltic Sea region (hereafter referred to as the Baltic Sea system; labeled "B" in Fig. \ref{fig:synoptic-chart}c,d).
All three convective systems have their main outflow at upper tropospheric levels \citep[about 200-300 hPa; see also ][]{Grootetal2023,grootthesis}, will be shown to interact with the PV gradient associated with the jet stream, and their variability plays a prominent role in the evolution of ensemble spread.
The nature of their interaction and the mechanisms that govern spread growth will be the focus of Sect.~\ref{alps-section}--\ref{36hoursdiscussion}. 
%In this work we investigate the significance of the intensity of mesoscale convection in the warm air masses for the jet stream location on the flanks of (and downstream of) this ridge, which itself lies over the British Islands and North Sea. On the first day, the baroclinic zone triggers convective systems over the Alps and Germany (Fig. \ref{fig:convective_systems}a, b). A shortwave through triggers convective systems over Germany on the second day. All of the convective systems move northward (Fig. \ref{fig:convective_systems}). The northernmost systems propagate to Denmark, Sweden and the surrounding seas on the second day (Fig. \ref{fig:convective_systems}d). 
%
%Convective systems can be expected to have their main direct impact on the spread evolution in very similar ensemble error-growth experiments during the first 2--3 days \citep{Selzetal2021,Baumgart2019}. We therefore limit the scope of our analysis to the first 3 days of our experiment starting on 10 June, 12UTC.
We limit our analysis to the first 3 days of our experiment, because the main direct impact of the divergent outflow on the spread evolution occurs during the first 2--3 days in comparable experimental configurations \citep{Baumgart2019,Selzetal2021}.
\subsection{Characteristics of the convective systems}
\label{sec:precip-systems}
\begin{figure}[t!]
    \centering
    \includegraphics[width=155mm]{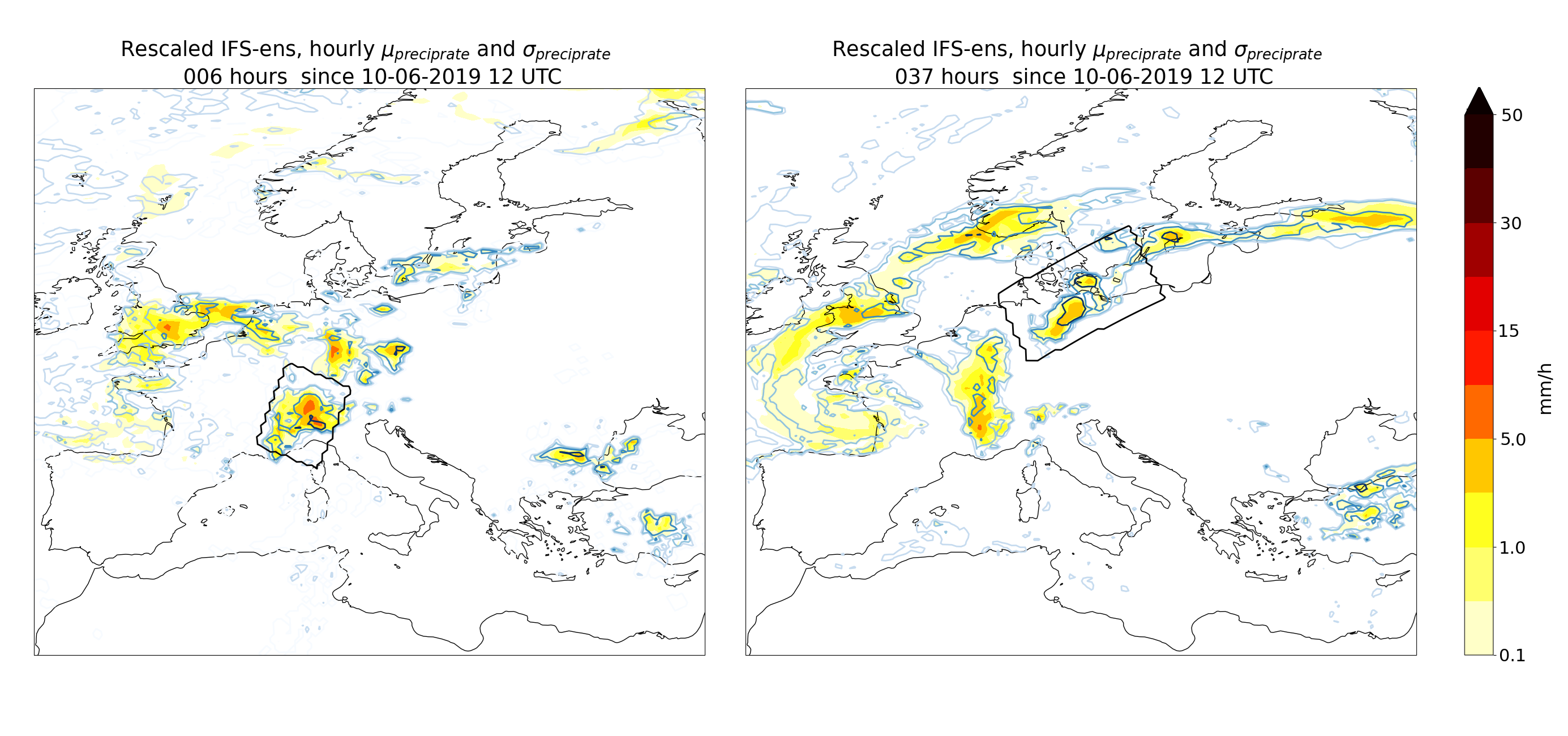}%\includegraphics[width=155mm]{convective-system-outlines/test_DAY2_L37.png}\\
    \caption{Mean precipitation rates (fill) and spread in mean precipitation rate (isolines) at approximately the time of maximum intensity for (left) the Alps System and (right) the Baltic Sea system. Spread is contoured at grid values of 0 (i.e. precipitation occurs in some ensemble member(s)), 0.01, 0.05, 0.25 and 1 mm/h. % one hour after the initial time step over which mass divergence and precipitation rate are integrated over 
    The masks (black outline) defining the extent of each convective system is also visualised at this (approximate) time of maximum intensity. %and one hour before the final time step of integration (right, panels b and d). Top: Alps system (day 1; stationary); %Middle: Northern Germany system (day 1; moving); 
    %Bottom: Baltic Sea system (day 2; moving)
    }
    \label{fig:convective_systems}
\end{figure}

As introduced in Sect.~\ref{methods}, we describe characteristics of the convective systems by defining a region for the respective system, over which quantities are averaged.
%We fit an averaging  volume over which he outflow rate of each convective system and the corresponding instantaneous precipitation rate is computed%(over the corresponding area), following \cite{Grootetal2023,Groot_Tost_2022b}. 
The horizontal extent of these regions are displayed in Fig. \ref{fig:convective_systems}.
For the Alps System, the region remains fixed. % and moving with the convective system with a fixed velocity for the Baltic Sea System (to the northeast)
For the Baltic Sea system, the averaging region moves with the convective system with a constant northeastward translation speed.
A representative relative position of the region's boundary with respect to the embedded precipitating area is depicted in Fig. \ref{fig:convective_systems}b.
For the Germany system, our analysis presented in Sect.~\ref{sec:results} below does not indicate that convective variability directly projects onto variability of the strong PV gradient by variability of the convective outflow. 
%(while the PV gradient also remains disconnected from its main PV spread feature throughout).
We therefore refrain from a more detailed quantitative analysis using convection characteristics and,
consequently, herewith we omit the definition of an averaging region for that system. 

Based on our averaging regions, the Alps system in our ICON experiments exhibits a mean precipitation rate of about 1.0 mm/h during its initial stage, first decaying to less than 0.7 mm/h after about 13\,h into the experiment, followed by further decay.
The Alps system accumulates 15 mm of precipitation, of which 9.4 mm falls between 2 and 13h lead time.
% simulation time, and eventually 0.2-0.3 mm/h. %Another convective system occurs further north over northern Germany on the first day. However, since the main interaction between the convection and PV gradient seems to be associated with the Alps system (located closer to the gradient), the extent of the northern German system needs not to be defined.
%The Baltic Sea system in our ICON experiment ... SOMETHING SEEMS TO MISS HERE ABOUT THE B SYSTEM.
The key event on day 2 is the development of the Baltic Sea system (Fig. \ref{fig:convective_systems}). Shortly after its initiation over central Germany (after about 31\,h), its precipitation rate doubles between 34\,h--36\,h over northern Germany. While moving to the Baltic Sea,
% (with fixed northeastward translation rate of integration volume), 
the mean precipitation rate fluctuates between 0.22 and 0.42 mm/h. Despite the lower intensity, appreciable precipitation (3.6 mm accumulation on average) occurs over a notably larger area than covered by the Alps system.  

%The ensemble variance in the precipitation rate tends to increase with time, in line with intuition. However, these tendencies do not manifest uniformly in space and time. 
%The ensemble spread of the accumulation over the Alps System's box within the ensemble is 0.04 mm. The corresponding numbers for the Baltic Sea system are 3.6 mm and 0.10 mm. 
% Table to display those values + check of the number: it should be 0.4 rather than 0.04!
Consistent with very small initial condition uncertainty, and the use of a deterministic convection scheme, the spread in the precipitation associated with the two systems is very small: 0.4\% and 3\% relative to the mean precipitation, respectively (Table \ref{rainfallchar}).
Despite this small relative spread, both systems can be identified as distinct features in the PV spread at the time of their occurrence and PV-spread tendencies can be clearly associated with these systems (see below).
Furthermore, we note that the relative spread of the Baltic Sea system, which occurs on day 2 in our experiments, is an order of magnitude larger than that of the Alps system, which occurs on day 1.
This order-of-magnitude difference is consistent with exponential initial spread growth, i.e., while spread is still small and far from saturation, and very short spread-doubling times in convective regions
%While convective scale errors are known to grow rapidly 
\citep[order of hours, e.g.][]{hohenegger2007,Selz2015a,Weyn_Durran_2017,Groot_Tost_2022}. It is illustrative for the tiny magnitude of our (convective) forecast spread, despite the lead time of \~36~h.  
%and the second convective system we mainly study lags the first by a day, the relative spread is much smaller than the mean intensity of both systems: %, despite their time lag: on the order of a percent (Table \ref{rainfallchar}).
%
%A convective system further north over Germany, downstream of the Alps System on the first simulation day, also produces distinct spread. However, despite its ability to produce spread growth within range of the PV-gradient, the produced couples weakly (if at all) to a straight PV-gradient. Therefore, a deep quantitative analysis would gain little insight (within ICON's EU nest). % \color{red}consider adding a plot of time evolution of respective precipitation rates\color{black} 
\begin{table}[tb]
\caption{Accumulation and spread in rainfall in two convective systems.}
\label{rainfallchar}
\begin{tabular}{|l|l|l|}
\hline
\textbf{Name}                                       & \textbf{Alps System} & \textbf{Baltic Sea System}  \\ \hline
$T_{start}$ (h)                           & 2           & 31                \\ \hline
$T_{end}$ (h)                             & 13          & 46                \\ \hline
Duration (h)                               & 11          & 15               \\ \hline
Accumulated rainfall (mm)                  & 9.4         & 3.6              \\ \hline
Ensemble spread accumulated rainfall (mm)  & 0.04         & 0.10             \\ \hline
Dimensionless ensemble spread rainfall (-) & 0.004        & 0.03            \\ \hline
\end{tabular}
\end{table}
\newline
% visible, for instance after +36 near 54 $^{\circ}$N, 10 $^{\circ}$E. Their visibility, with low amplitude, suggests that deep convective cells are underrepresented considerably in this used configuration, as anticipated upon. %Nevertheless, weak variability over Southern Sweden might be present after 42 hours, associated with the convective systems outlined in the previous Section.
The spread of convective outflow is very similar to that of precipitation (as in \cite{Grootetal2023}; here further demonstrated by results shown in Appendix \ref{prec_var}).
%Following the findings of \cite{Grootetal2023}, we can think of the precipitation rate variability and outflow rate variability as being highly correlated for our setup. This assumption is supported strongly by the findings of Appendix \ref{prec_var}. 
Our analysis in Section \ref{sec:results} therefore focuses on the relation between outflow and PV variability, without further explicit consideration of precipitation variability. %, as detailed on in Section \ref{sec:results}. The perturbations are expected, in relation to the divergent outflows in the upper troposphere and their associated negative PV. \newline
%The volumes displayed in Figure \ref{fig:convective_systems} are used for the integration of mass divergence (box volume) and precipitation rate (areas) in the ensemble sensitivity analysis, as part of the following Section \label{sec:results}. 
%\begin{itemize}
%    \item This section should define the precip %signals of the three investigated convective systems
%    \item Mention at least that the convective system over the Baltic Sea strengthens substantially at the 34-37 hour simulation time and by how much it does. 
%    \item Maybe mentioned thing in the next section!
%\end{itemize}
%\section{Results}
%\label{sec:results}
%
%
%
\subsection{Overview of spread evolution}
\label{spreadevo}
\begin{figure}[tb]
    \centering
    \includegraphics[width=160mm]{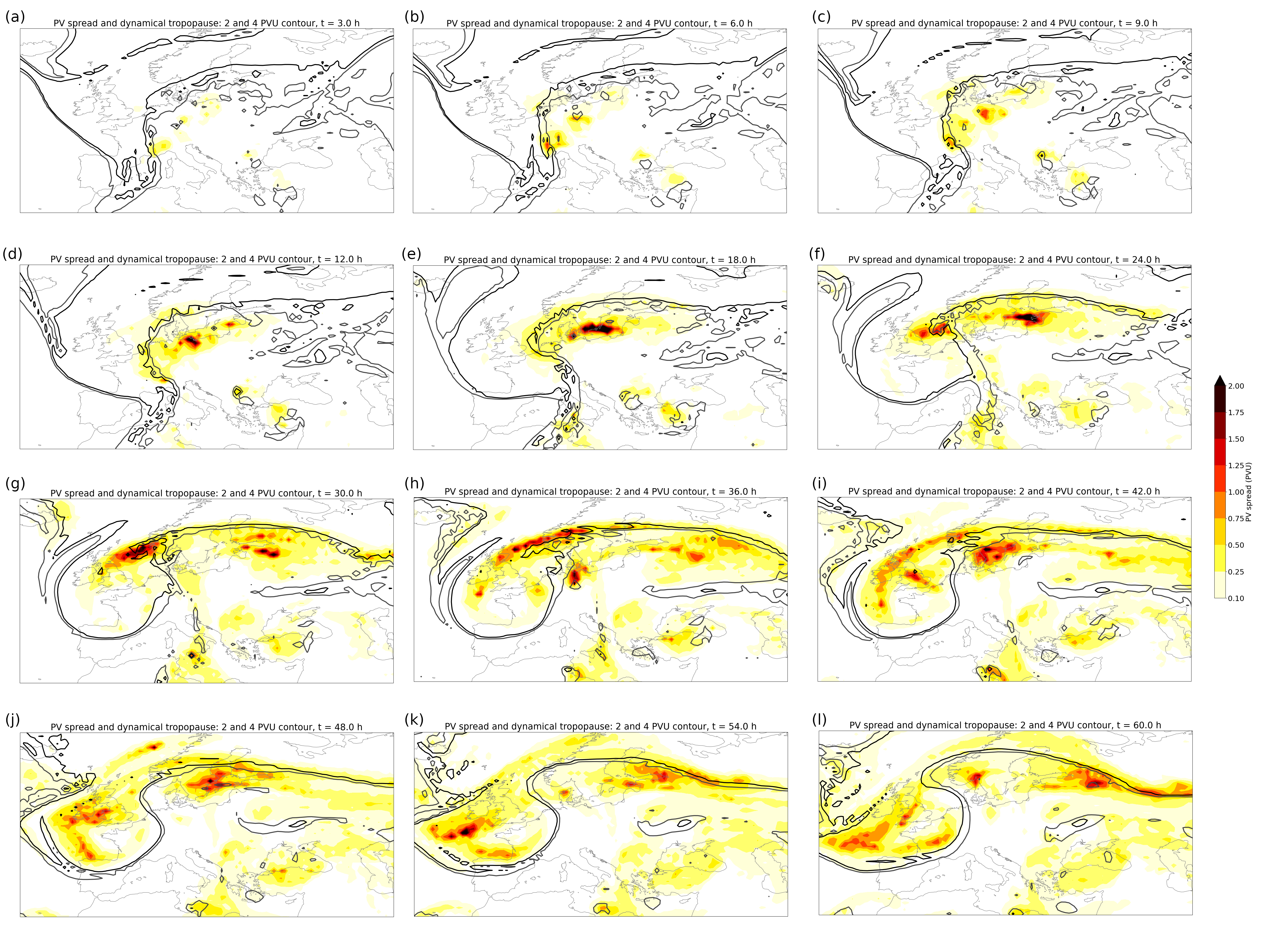}
    \caption{Evolution of 2 and 4 PVU contours at 330 K (ensemble mean) with PV-spread at 327.5-335 K superposed (colour fill) at 3 hour intervals (initially) and 6 hour intervals (beyond 12~h).% Note that the contour values of ensemble standard deviation increases after 30 hours. 
    }
    \label{fig:meanPVpanel}
\end{figure}

%--- MR suggestion for  revised version: ---
%
We provide an overview of the spread evolution in our spread-growth ensemble experiment on isentropic levels intersecting the dynamical tropopause in Fig.~\ref{fig:meanPVpanel}. This figure will further serve as reference when discussing our process-based analysis of the spread evolution in Sect.~\ref{sec:results}.

Early during the experiment (until 9\,h, Figs.~\ref{fig:meanPVpanel}a-c), spread is predominantly associated with the Alps and Germany system.
Subsequently, spread over the Germany system amplifies strongly but remains localized in the vicinity of the system. %Beyond the most active convective systems and away from the adjacent PV gradient, spread is hardly amplified on the first day.
%The dominant spread production in the first hours (about +3\,h lead time) with downstream impact occurs over southern France, adjacent to the convection over the Alps. This region of s
Spread associated with the Alps system seemingly extends downstream (northward) along the strong PV gradient associated with the upstream trough and eventually amplifies during the cyclonic wrap up of the trough (Figs.~\ref{fig:meanPVpanel}d-f).
At 24\,h, spread associated with the trough and with the Germany system constitute prominent spread maxima (Figs.~\ref{fig:meanPVpanel}d-f).
On day 2 of our experiment (Figs.~\ref{fig:meanPVpanel}g-j), the spread associated with the Germany system is being advected downstream without prominent further amplification, while the spread associated with the trough wraps up cyclonically within the trough and exhibits limited impact on the downstream ridge. 

Associated with the Baltic Sea system, a new local spread maximum emerges around 36\,h over northern Germany (Fig.~\ref{fig:meanPVpanel}h), subsequently moving over the Baltic Sea (42--48\,h, Fig.~\ref{fig:meanPVpanel}i-j).
This spread maximum apparently couples with the strong PV gradient and amplifies while propagating downstream (54--60\,h, Fig.~\ref{fig:meanPVpanel}k-l).
Late in our experiment (exemplified at 60\,h in Fig.~\ref{fig:meanPVpanel}l), spread is predominantly located in the vicinity of the strong midlatitude PV gradient.
For completeness, we note that spread has developed also in the southeast of our domain, originating from convection in that region (as indicated below in Fig.~\ref{fig:meanPVpanel}). 
This spread is of lower amplitude than that over western and northern Europe (at least on the examined isentropic levels) and does not interact with the strong midlatitude PV gradient in any obvious way. Therefore, 
%this spread source is beyond the scope of our work.
spread evolution in that region of our domain is not relevant to this study. 

\color{black}%Furthermore, a small-scale through forms a PV-filament over Central-Europe (about 1 PVU only) as the blob of \> 10 PVU propagates northward. Ahead of this through, conditions in the warm air mass are getting increasingly favorable for deep convective systems. Subsequently, the threshold of each isoline of PV-spread is raised. \newline
%Note that as a feature of large PV-spread moves through the warm air from Northern Germany into Southern Sweden (+36, +42 hours), the filament of largest PV-spread over Central Scandinavia appears to re-locate at about 100-300 km south of its original location. The evolution associated with re-location of the PV-spread and PV-gradient is closely investigated in one of the next sections: Section \ref{sec:balticsystem}.  
%\begin{figure}
 %   \ContinuedFloat
  %  \caption{Continuation of the Figure on the previous page. %Mean potential vorticity at 250 hPa (colors) and ensemble standard deviation in potential vorticity (grey to black) in isolines. Every 3 hour intervial is initially displayed and later on every 6 hour interval (with an extended map in the last panels).
  %  }
   % \label{fig:meanPVpanel}
%\end{figure}
\section{Results: analysis of spread dynamics}
\label{sec:results}
%\textcolor{red}{The titles should contain the names of the systems. Should make clear again that our interest is in the 3 stage sequence when we look at the spread tendencies and ESA. LW tendencies will need some extra intro.}
In Sect.~\ref{alps-section}-\ref{36hoursdiscussion} we analyse the PV-spread-tendencies associated with the three convective systems introduced in Sect. \ref{sec:precip-systems}. In addition, we employ ensemble sensitivity to link convective variability with ensuing PV variability along its near-tropopause gradients. Both analyses have been carried out over the full European nest (13 km grid spacing) and are furthermore regridded to 0.75 degrees grid spacing for the analysis (approximately the effective resolution). Our focus with these analyses will be on the three-stage sequence of spread-growth processes as identified by \cite{Baumgart2019,Selzetal2021} and described in the introduction. In addition to the three dominant processes emphasized in these previous studies, i.e., the tendencies from the deep-convective scheme (stage 1) and their transition to divergent mode (mostly stage 2) and later rotational mode (mostly: stage 3), we will find a substantial contribution of longwave radiative tendencies occurring in close relation to convective activity. In Sect.~\ref{longwave} we thus examine this relation in some more detail. It turns out that tendencies due to grid-scale precipitation and shortwave radiation are typically small and tend to reduce spread (if active at all).
Throughout Sect. \ref{alps-section}-\ref{36hoursdiscussion}, we present three hour averages of selected intervals, which are typically representative of nearby intervals and a smooth evolution (unless it is indicated they are not).

\subsection{Alps convective system}
\label{alps-section}
\begin{figure}[htb!]
    \centering
    \includegraphics[width=160mm]{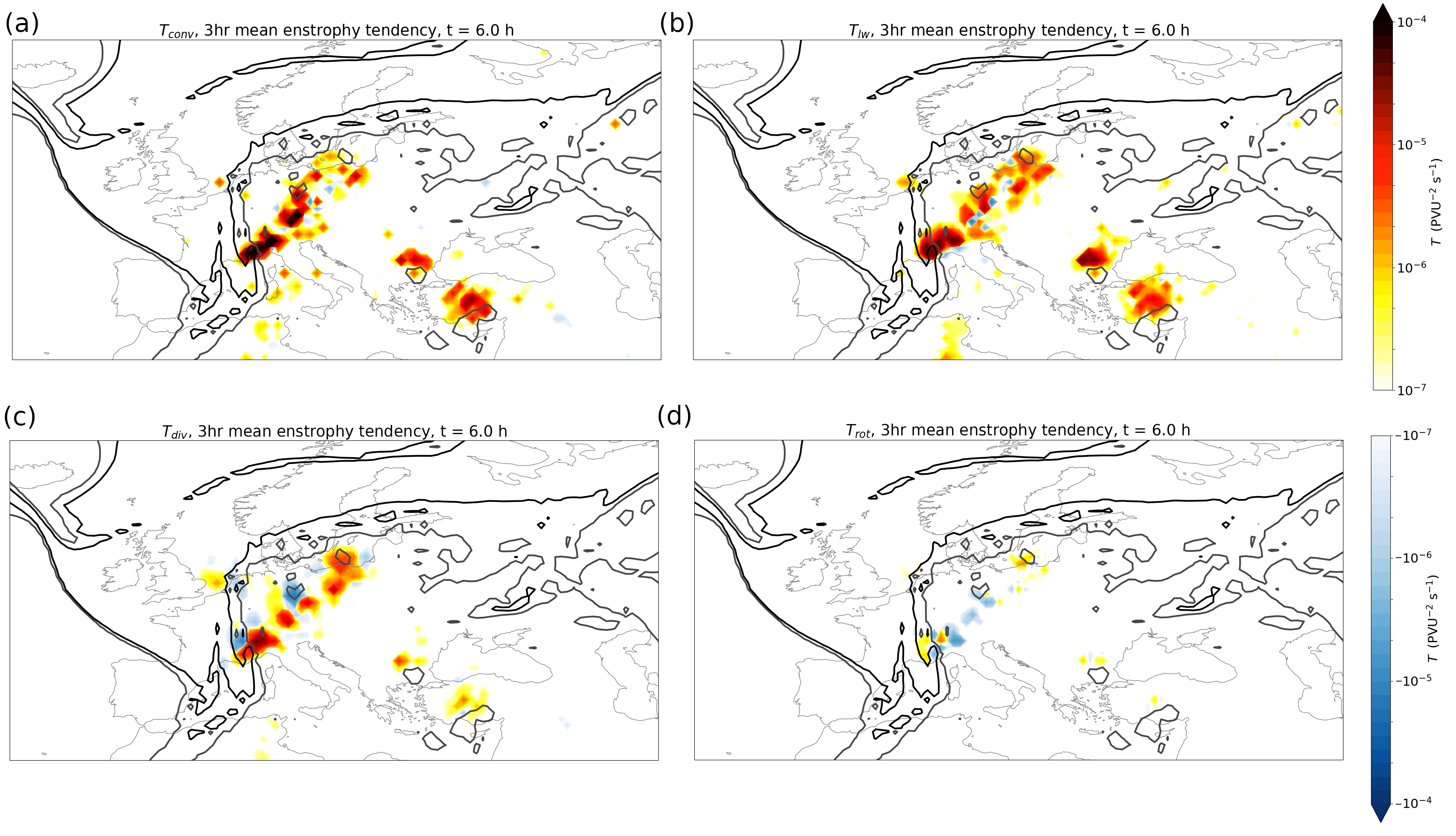}\\
    \caption{Three hour mean of the potential enstrophy tendency for two different processes at $\theta$ of 327.5-335 K, $t=6$~h:  heating by the deep convection parameterisation (a), the heating by longwave radiation parameterisation (b), the differential PV-advection resulting from differences in divergent winds (c) and similarly for the rotational winds (d). Red colours indicate spread growth, whereas blue colors indicate spread reduction for each term. %Green isolines highlight regions of enhanced PV-spread at 330 K. 
    For this level, the dynamical tropopause (2PVU) and 4 PVU contour are also shown in dark grey/black.}
    \label{fig:PVdiags-Alps-system}
\end{figure}

The first spatially coherent signal of spread-growth rates in the vicinity of the Alps system occurs after 3h-6h (Fig. \ref{fig:PVdiags-Alps-system}; note that we illustrate 3h average growth rates for optimal clarity). Growth rates due to the deep-convective scheme dominate in the vicinity of the Alps system at this early stage of spread growth (Fig. \ref{fig:PVdiags-Alps-system}a), in close proximity to the jet stream (2 and 4 PVU contours) and matching with the area of PV spread (cf.~Fig. \ref{fig:meanPVpanel}). Divergent spread growth also occurs in the vicinity of the Alps system, but with much smaller amplitude (Fig. \ref{fig:PVdiags-Alps-system}c). Nonlinear spread growth is negligible (Fig. \ref{fig:PVdiags-Alps-system}d). The early-stage spread growth near the Alps system is thus confirming dominance of a first spread-growth stage, in line with \cite{Baumgart2019,Selzetal2021} (and of \cite{Zhang2007}).
%The pattern of the convective tendencies matches the pattern of ensemble spread at this early stage very well (cf.~Fig. \ref{fig:meanPVpanel}), confirming dominance of stage 1 in \citep{Zhang2007,Baumgart2019, Selzetal2021}. % This dominance of spread growth by the convective tendencies confirms that the convective variability is of primary importance during this early stage for PV-ensemble spread \citep{Baumgart2019, Selzetal2021}. %, which is expected at short lead times based on earlier analyses of hemispheric averages in ensembles with a stochastic convection scheme \citep{Baumgart2019, Selzetal2021}. %, given the early stage of the simulation and, correspondingly, comparatively low PV-spread.% in PV.% (in the upper troposphere and near the tropopause). The combination of Figures \ref{fig:PVdiags-Alps-system} and \ref{fig:convective_systems} (top panel) therefore suggests local structural upper tropospheric PV-modification over the area of mesoscale convection in the Alps. 
 
\begin{figure}[b!]
    \centering
    \includegraphics[width=160mm]{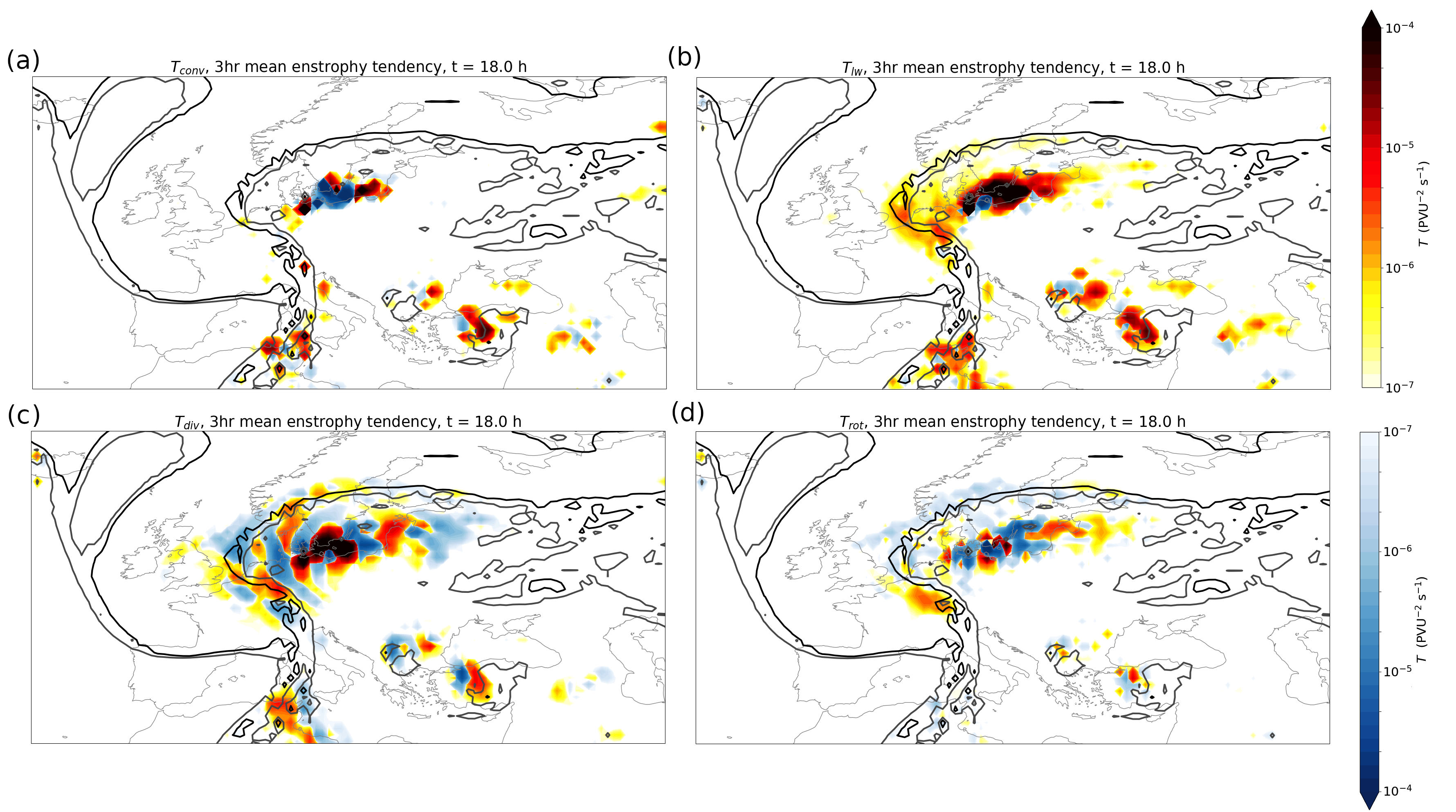}
    \caption{As in Fig. \ref{fig:PVdiags-Alps-system}: three hour mean of the potential enstrophy tendency for four different processes at $\theta$ of 327.5-335 K, $t=18$~h.}
    \label{fig:18hr-tendencies}
\end{figure}

We further note that longwave radiative tendencies make a notable contribution to spread growth near the Alps system (Fig.~\ref{fig:PVdiags-Alps-system}b). The magnitude of the associated growth rates is larger than that of the divergent tendencies and somewhat smaller than that of the deep-convective scheme, while all three overlap spatially. On the broader scale, the longwave radiative tendencies exhibit a remarkable spatial consistency with the convective tendencies (cf.~ Fig.~\ref{fig:PVdiags-Alps-system}a,b) and, overall, are of similar magnitude. This general spatial consistency persists over the next 15 hours and can be observed also at 15~h-18~h (Fig.~\ref{fig:18hr-tendencies}a,b).
%The second most important contribution to spread growth at this early time is by the longwave-radiative tendencies, which are spatially coincide with the convective tendencies (Fig. \ref{fig:PVdiags-Alps-system}b). 
The vertical structure of their close relationship will be discussed in more detail in Sect. \ref{longwave} for one convective system.
%The other panels of Figure \ref{fig:PVdiags-Alps-system} reveal that strong spread tendencies are not only associated with deep convection (CONV) (panel a; initially dominant), but also strongly with longwave radiation (LW), coming second (panel b), and thirdly with divergent winds (DIV; panel c).
%The divergent tendencies are spatially also related to convection, but are still weaker than the convective and longwave-radiative tendencies (Fig. \ref{fig:PVdiags-Alps-system}c).
%Maximum values are found over the Alps system, where intense mesoscale convection is closest to the large PV gradient associated with the jet stream.
%These divergent wind tendencies either are either caused by areal expansion/shrinkage of pre-existing PV-spread through divergent/ convergent advection of pre-existing enstrophy or by the spatial gradients in the local enstrophy fluxes. 
%Therefore, of these three terms, only the DIV term allows to redistribute spread spatially. 
%The general patterns of the convective, longwave-radiative, and divergent spread tendencies
%, initiated by the deep convection parameterisation, longwave radiation and divergent winds are 
%persist over the next 15 hours (cf.~\ref{fig:18hr-tendencies}a-c).

It is difficult to discern a period when spread growth associated with the Alps system is dominated by divergent tendencies, i.e., it is difficult to identify the dominance of the second stage of the three-stage models of \citep{Zhang2007,Baumgart2019, Selzetal2021}. After dominant early-stage spread growth by deep convective tendencies, the divergent tendencies exhibit a rather complex pattern of alternating positive and negative values (exemplified at 15h-18h in Fig.~\ref{fig:18hr-tendencies}c), in particular in the region of the spread associated with the Alps system located on the northeastern flank of the upstream trough (cf.~Fig.\ref{fig:meanPVpanel}a-e). Visual inspection of divergent spread tendencies at intermediate times strongly indicates that the alternating pattern is associated with inertia-gravity waves emanating from the Alps system and the German system (not shown), which fills the region of this ridge over Central Europe. The wave-like pattern of the divergent tendencies prohibits a straight-forward interpretation of its impact on the spread associated with the Alps system and the PV-gradient in its proximity. Regardless, it is not obvious from visual inspection whether the positive values of divergent tendencies on the north-eastern flank or rotational and longwave radiative tendencies net dominate this region. 

\begin{figure}
    \centering
    \includegraphics[width=120mm]{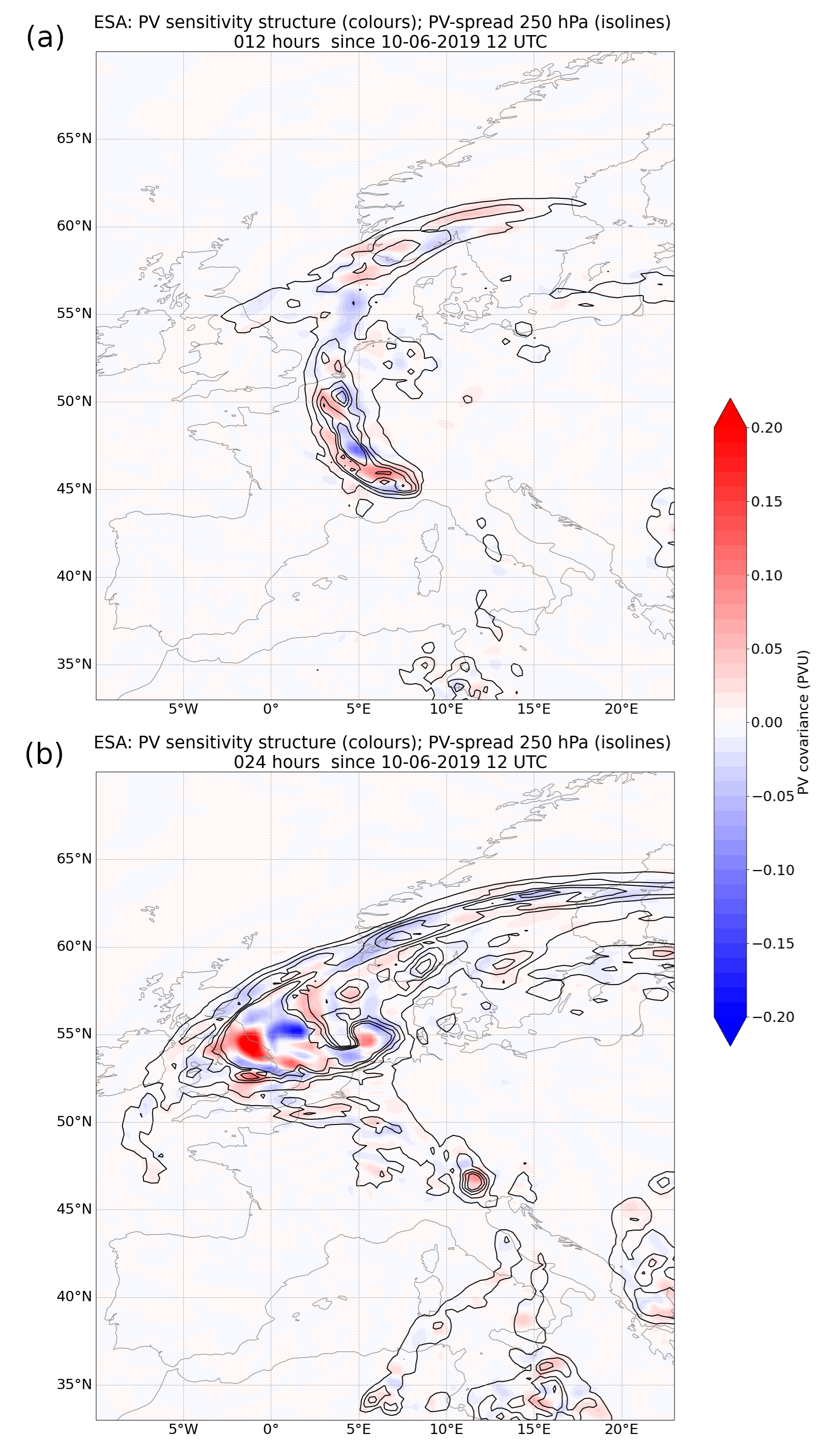}%\\\includegraphics[width=120mm]{esa-results/ESA-alps-system/ESA_PV_signal_Alps_System_t2_to_t5_div_hPa_PV_0048.png}

    \caption{Potential vorticity covariance at 250 hPa within the ensemble associated with enhanced mass divergence in the upper troposphere over the convective system in the Alps between 2~h and 5~h simulation time. Top: 12~h; bottom: 24~h. Isolines indicate regions of enhanced $\sigma_{PV}$ within the ensemble (isolines at 0.06 PVU intervals between 0.06 and 0.30 PVU).  }
    \label{fig:ESA_alps_system}
\end{figure}

It is noteworthy that longwave tendencies in the spread region associated with the Alps system does not coincide spatially with deep convective tendencies at 15h-18h (cf.~Fig.\ref{fig:ESA_alps_system}a,b). Visual inspection of intermediate times (not shown) reveals that this local maximum of longwave tendencies can be traced back to % originate from over
the Alps system. Apparently, longwave-radiative tendencies outlive convective tendencies there, after the convective system has ceased. In the rest of the domain, longwave-radiative tendencies spatially coincide with convective tendencies as at 6h, but longwave tendencies are now mostly of larger magnitude than the deep convective tendencies.
%The nonlinear (rotational) tendencies, which are most directly associated with the existing spread, make a very minor contribution at this early stage (Fig. \ref{fig:PVdiags-Alps-system}d).
%, although the northeasterly convection intensifies and the system over the Alps looses intensity beyond 12 hours. 
%\newline 
%Furthermore, between about 6 and about 12 hours, a transition leading to more widespread spread tendencies, often with dipole signature, takes place. The spread tendencies beyond this transition are analysed in the final paragraphs of the current Subsection. The spatial distribution of CONV, LW and DIV tendencies are highly related in the first hours, where co-location of CONV and DIV spread sources is consistent with the main hypothesis of this work. We investigate the relation between CONV and LW further in the last subsection of this section. We now switch our focus to the ensemble sensitivity analysis and then further interpret findings from both methods chronologically. 
%

To further explore the relation between the variability of convective outflow with PV spread we employ ensemble sensitivity analysis. This analysis shows that the strength of the divergence over the Alps system and PV values along the adjacent large PV gradient over eastern France are associated with each other (Fig.~\ref{fig:ESA_alps_system}a). The analysis thus directly quantifies the link between the variability in convective activity of the Alps system and nearby jet variability. However, since dipole patterns are seen and absolute linear correlations are not uniform, there is no suggestion of a direct linear link between the two, indicating no %. Increased outflow rates of the Alps System are thus not 
uniform connection between outflow rates of the Alps System and adjacent %ly associated with 
reduced PV values. Such a uniform area of negative correlations would be expected if stronger outflow were associated with a stronger displacement of PV contours from the ridge towards the trough, thereby leading to reduction of PV in the region of displacement.

%Consistent with the indications of the sensitivity analysis, the divergent spread tendencies indicate that a substantial interaction of the outflow with the jet occurs only near the tip of the breaking trough. These tendencies do not indicate interaction throughout the broader region of enhanced spread, from the Alps to the North Sea (cf.~Fig.~\ref{fig:PVdiags-Alps-system}c and Fig.~\ref{fig:ESA_alps_system}a).
%The alternating positive and negative correlations of the sensitivity analysis are indicative of a downstream propagation of the highly localised perturbation of the jet by the Alps system. 
%Despite the double dipole pattern, the convective outflow rate has a significant correlation with a reduction of the mean PV over the spread maximum in eastern France - at least after 12 hours at the expected time lag, 
%although barely reaching the significance threshold. %However, clearly, the hypothesised simple outflow-jet interaction is not supported in convective event. 
%The ensemble sensitivity analysis thus confirms that variability in the convective activity of the Alps system generally translates to spread along the adjacent jet.
%The rich regional substructure, however, seemingly related to the highly localized impact of the Alps system does not corroborate our hypothesis presented in the introduction.
%\newline

Our interpretation of the alternating positive and negative correlations of the sensitivity analysis is that they indicate a downstream propagation of a more localised perturbation of the jet by the Alps system (as indicated in Fig.~ \ref{fig:ESA_alps_system}a-c). Despite this double dipole pattern, the convective outflow rate has a significant correlation with a reduction of the mean PV over the spread maximum in eastern France - at least after 12 hours at the expected time lag, although barely reaching the significance threshold. The ensemble sensitivity analysis thus confirms that variability in the convective activity of the Alps system generally translates to spread along the adjacent jet. The rich regional substructure, however, seemingly related to a more localized impact of the Alps system, does not support the notion that divergent tendencies dominate spread growth associated with the Alps system for a certain period of time, i.e., there is no clear evidence for the occurrence of a distinct second stage of three-stage spread growth for the Alps system.

The third stage of spread growth (according to \cite{Baumgart2019,Selzetal2021}) is dominated by rotational tendencies. At 15h-18h, the rotational tendencies have clearly amplified in absolute values since at 3h-6h. The northeastern flank of the trough is a region of spatially coherent positive tendencies, clearly linked to the local spread maximum initiated earlier by the then occurring Alps system (Fig.~\ref{fig:18hr-tendencies}d), although the convection itself has now mostly deactivated. This coherent spread growth indicates the highly nonlinear evolution of the trough, i.e., wave breaking, because nonlinearity in the underlying dynamics is required for spread amplification by rotational tendencies \citep{Baumgart2018}. Despite spatial coherency, the rotational tendencies are very localised and, hence, cannot dominate spread growth on the flank of the trough at 15h-18h. Subsequently, the trough completes its cyclonic breaking and spread amplifies within and spreads across the broken trough (Fig.~\ref{fig:meanPVpanel}e-j). The tendencies depicted at 33h-36h in Fig.~\ref{fig:PVdiags_36h} are representative of the relative importance of individual contributions to spread growth associated with the trough. Overall, the tendencies exhibit complex patterns, but it is clear that the rotational tendencies do not dominate spread growth in the trough region. Due to the dominant contributions by longwave tendencies at this time, which cannot be traced back to the Alps system, it is clear that the further generation in the trough has little origin (if any at all) in the convective uncertainty associated with the Alps system.

In summary, the evolution of spread associated with Alps systems only partly exhibits signatures of the hypothesized sequence of events identified from diagnostics after averaging \citep[by][]{Baumgart2019, Selzetal2021}. 
Convective tendencies dominate initial spread growth (an indication of stage 1). Indicating a presumed transition to stage 2, divergent tendencies generate PV spread along the strong PV gradient adjacent to the Alps system. Rotational 
%Whereas convective tendencies dominate initial error-growth (an indication of stage 1), divergent tendencies amplify spread near the Alps system early on (indicating a transition from stage 1 to stage 2 of \cite{Zhang2007,Baumgart2019}) and generate PV spread along the adjacent strong PV gradient. 
%Then rotational 
tendencies further amplify the existing spread by nonlinear dynamics along the PV gradient, indicating the onset of stage 3.
Despite these consistencies with the conceptual sequence, subsequently, divergent and rotational tendencies do nevertheless not play a sustained, dominant role. Hence, pronounced and distinctive fingerprints of stages 2 and 3 cannot be identified. Arguably, nonlinear spread growth did not have the time to become a dominant mechanism, because of the cyclonic breaking of the upstream trough. This breaking effectively limited downstream dispersion and associated further amplification of spread by rotational tendencies. In addition, longwave-radiative tendencies made a substantial contribution to the spread evolution throughout the considered period, further complicating the identification of a three-stage sequence. Finally, significant interaction between the Alps System and the PV-gradient is suggested by both of our analyses methods.

\color{black}
\subsection{The Germany system and its spread evolution}
\label{day2results}

%\comment{To be inserted in this subsect.: Further north, in the building ridge, convective tendencies are of mixed sign, as opposed to a predominantly positive contribution by +6 h (Fig.~\ref{fig:18hr-tendencies}a vs. \ref{fig:PVdiags-Alps-system}a).}

%At the end of day 1, two distinct regions of enhanced spread exist (Fig.~\ref{fig:meanPVpanel}f).
%One region is located over the North Sea and associated with the cyclonically breaking trough, partly originating from the Alps system as described in the previous subsection. The spread hardly interacts with any spread further downstream, and its further evolution is beyond the scope of this work.
%The other region is located over the Baltic Sea, originating and moving with the convective system that developed over Germany during day 1. Here, we analyse their characteristics and then we investigate whether these match with stages 2 and 3 of the conceptual model for error-growth, as described by \cite{Zhang2007,Baumgart2019}, before moving on to the phase where stage 1 can reoccur with the initiation of the Baltic Sea System.
%During day 2, the spread associated with the breaking trough is mostly confined to the trough itself, with an indication of some remnant spread over Norway at 36-48h (Fig.~\ref{fig:meanPVpanel}f-j). 

Spread associated with the Germany convective system develops early in the experiment over Germany (at 3h-6h and 6h-9h, Fig.~\ref{fig:meanPVpanel}a-c, cf.~Fig.~\ref{fig:convective_systems}a).
Subsequently, moving northeastward within the ridge over central Europe (closer to the PV gradient), this spread region constitutes the distinct maximum of the spread distribution until 24h-27h (Fig.~\ref{fig:meanPVpanel}c-f). Moving along the zonally oriented, strong PV gradient over northeast Europe (30-42h, Fig.~\ref{fig:meanPVpanel}g,i), the maximum tends to weaken and there is no clear net amplification of spread when interacting with the jet. At the end of day 2, spread associated with the Germany system has essentially moved out of the domain under consideration.

%The spread associated with the German convective system of day 1 moves downstream and closer to the strong PV gradient associated with the jet (Fig.~\ref{fig:meanPVpanel}f-j).
%There is no clear net amplification of spread when interacting with the jet.
%At the end of day 2, this enhanced region of spread has moved out of our integration domain.
%Note that the region downstream surrounding the Baltic Sea exhibits very little spread at 30-36h, just like its immediate downstream region in the Baltic States after 42-48h (Fig.~\ref{fig:meanPVpanel}j).

Similarly to the Alps system, initial spread growth over the Germany system is dominated by deep convective tendencies, accompanied by a secondary contribution of longwave radiative tendencies (Fig.~\ref{fig:PVdiags-Alps-system}a,b). The first, convective stage of spread growth is thus again evident. In contrast to the Alps system, divergent tendencies subsequently make a clear and strongly positive contribution to spread growth above the Germany system: While positive and negative values, supposedly associated with the inertia-gravity waves, fill most of the ridge over Europe, divergent tendencies remain strongly positive just above the Germany system for the remainder of day 1 (illustrated in Fig.~\ref{fig:18hr-tendencies}c at 15h-18h). Meanwhile, net amplification of spread by deep convective tendencies has clearly decreased over the system (Fig.~\ref{fig:18hr-tendencies}a).  The inertia-gravity wave pattern extends to the strong PV gradient associated with the jet (between the North Sea and the Baltic Sea; Fig.~\ref{fig:18hr-tendencies}c). At a fixed location, however, the sign of the tendency alternates with time (not shown). The characteristics of divergent tendencies remain qualitatively similar during day 2 (not shown), while the Germany system is moving somewhat closer to the strong PV gradient (exemplified at 33-36h in Fig.~\ref{fig:PVdiags_36h}c): The divergent tendency pattern is complex, with large positive values above the active convection of the system, but no persistent positive values along the strong PV gradient closest to the system, and hence, with no remote impact on that gradient.

In contrast to the divergent tendencies, rotational tendencies are positive along the PV gradient near the eastern part of the spread maximum associated with the Germany system on day 2 (Fig.~\ref{fig:PVdiags_36h}d, cf.~\ref{fig:meanPVpanel}h). This signal suggests that an interaction between the convective system and the jet occurs by the variability of the rotational wind that is in balance with the PV-spread generated by the convective system. As noted above, however, this interaction does not excite further substantial net amplification of spread along the jet within our integration domain. Before the associated spread pattern leaves the domain, the dominant spread source manifesting spread growth does not originate from rotational tendencies, which would represent stage 3 of the conceptual spread growth model \citep{Zhang2007,Baumgart2019}.

Similar to the Alps system, the longwave tendencies of the Germany system make an important contribution to spread growth, both on day 1 and day 2 (Figs.~\ref{fig:PVdiags-Alps-system} and \ref{fig:18hr-tendencies}b, and Fig.~\ref{fig:PVdiags_36h}b, respectively).
At 15h-18h, in particular, longwave tendencies are at least as strong as divergent tendencies (cf.~Fig.~\ref{fig:18hr-tendencies}b,c). Illustrated at 33h-36h, we note again that longwave tendencies overlap strongly with convective tendencies (cf.~ Fig.~\ref{fig:PVdiags_36h}a,b).

%A plausible explanation for this lack of interaction is that the convective system is situated too far away from the jet, which either reduces or negliges the remote impact on the jet.
%Spread features cannot reach the jet without shrinking by the mostly negative divergent tendencies directly along the PV gradient.
%In contrast, rotational tendencies are positive along the PV gradient near the eastern part of the spread maximum associated with the convective system over the Baltic Sea (Fig.~\ref{fig:PVdiags_36h}d, cf.~\ref{fig:meanPVpanel}h).
%This signal suggests that an interaction between the convective system and the jet rather occurs by the variability of the rotational wind that is in balance with the PV-spread generated by the convective system. 
%As noted above, however, this interaction does not excite further substantial net amplification of spread along the jet within our integration domain. 
%We can safely conclude from the lack of amplifying interactions at the PV-gradient that spread tendencies do not represent stages 2 and 3 of the conceptual error-growth model of \cite{Zhang2007,Baumgart2019} well, at least here.

In summary, we interpret the salient features of the spread evolution associated with the German convective system as follows. A first stage of spread growth is dominated by deep convective tendencies and a second stage by divergent tendencies. The second stage, however, is not characterized by an interaction of convective outflow with a strong PV gradient. A plausible explanation for this lack of interaction is that the convective system is still too far away from the jet. We suggest that the local spread growth by divergent tendencies above the convective system relates to the adjustment of convective outflow to a balanced anticyclone \citep{Bierdel2017,Bierdel2018}, which is thus more akin to a second stage of spread growth as suggested by \cite{Zhang2007} than to the sequence of \cite{Baumgart2019}. Along the same line, we interpret subsequent rotational tendencies along the strong PV gradient as an interaction of this anticyclone with the jet, which indicates the potential beginning of the third stage of spread growth dominated by rotational tendencies \citep[according to][]{Baumgart2019, Selzetal2021}. This stage, however, is not realized here, at least not within the domain of our experiment. We attribute the observed lack of subsequent spread growth by rotational tendencies to the rather straight orientation of the jet, which implies linear dynamics of perturbations on the jet. Nonlinear (Rossby wave) dynamics, however, are required for further spread amplification by rotational tendencies \citep{Baumgart2018}, and such a contrasting scenario will be exemplified below by the spread evolution downstream of the Baltic Sea system on day 3.

%In summary, we interpret the salient features of the spread  evolution associated with the German convective system as follows.
%There is a lack of downstream propagation - and potential amplification - of the spread associated with the breaking trough because wave breaking prohibits such propagation.
%The spread associated with the convective system over the Baltic Sea exhibits a downstream impact. 
%Interaction with the jet occurs by the rotational winds associated with this spread maximum, rather than by the divergent winds associated with the convective system.
%Along the jet there is little indication of further spread amplification.
%We attribute this lack of amplification to the rather straight orientation of the jet, which implies linear dynamics of perturbations on the jet. Nonlinear (Rossby wave) dynamics, however, are required for further spread amplification by rotational tendencies \citep{Baumgart2018}, which we here exemplify by the spread evolution on day 3 in the next subsection (Section \ref{36hoursdiscussion}).
%We further note that the spread associated with convection remote from the jet, i.e., in the south-eastern part of our domain, does not substantially amplify either during day 2.  

\subsection{The Baltic Sea System}
\label{36hoursdiscussion}
\begin{figure}[t!]
    \centering
    \includegraphics[width=160mm]{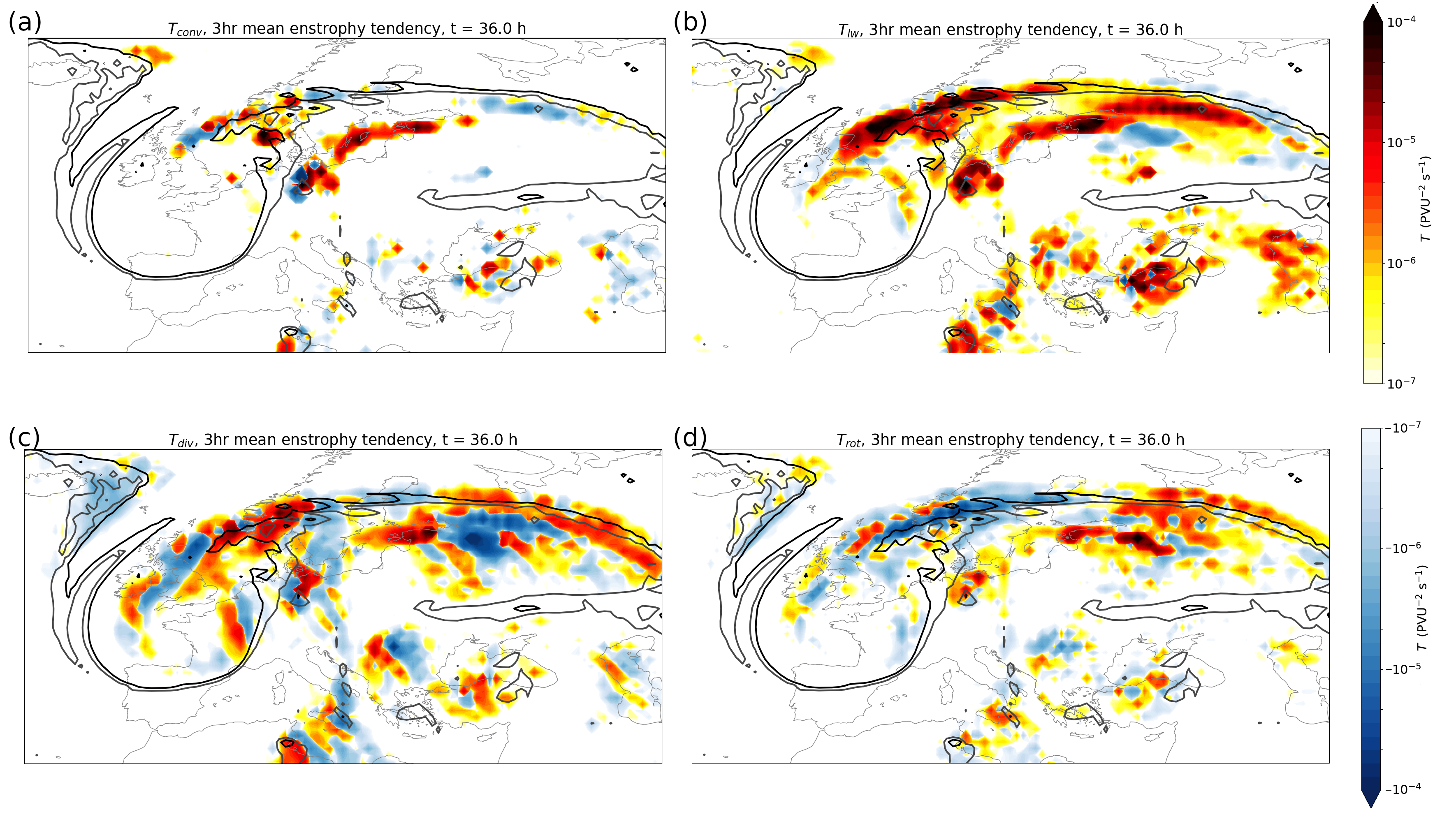}
    \caption{Same quantities as in Fig. \ref{fig:PVdiags-Alps-system}; three hour mean of the potential enstrophy tendency for four different processes at $\theta$ of 327.5-335 K, $t=36$~h.}
    \label{fig:PVdiags_36h}
\end{figure}

%Finally, the most prominent amplification of spread in our domain during day 2 is not related to the jet dynamics but is associated with the development of a new convective system over northern Germany.
%
Spread associated with the Baltic Sea system first occurs as a distinct maximum in the spread distribution around 36h over northern Germany (Fig.~\ref{fig:meanPVpanel}g; cf.~ Fig. \ref{fig:convective_systems}c). This spread maximum grows rapidly in the next 6h-12h and becomes dominant in the ridge over Europe, while the convective system is moving over the Baltic Sea and towards the strong PV gradient associated with the jet (Fig.~\ref{fig:meanPVpanel}h-i), cf.~ Fig. \ref{fig:convective_systems}d). Note that just before interaction with the strong PV gradient, spread is very limited along the PV gradient in the vicinity of the Baltic Sea system, as well as in the adjacent downstream region (Fig.~\ref{fig:meanPVpanel}i), because previous spread sources have left few traces (if any) of downstream impact here. After interaction, spread amplifies along the strong PV gradient (Fig.~\ref{fig:meanPVpanel}k-l). It is therefore feasible to examine the conceptual three-stage sequence of spread growth in the context of the spread evolution associated with Baltic Sea system, despite the relatively late initiation of that system on day 2 in our experiment. 
Just after the initiation of the Baltic Sea system, a first, convective stage of spread amplification occurs, evidenced by deep convective tendencies dominating over divergent and rotational tendencies in the vicinity of the system (shown at 33-36h in Fig.~\ref{fig:PVdiags_36h}a,c,d; cf.~Fig.~\ref{fig:convective_systems}c). Similar to the Alps and the Germany system, longwave tendencies make a further prominent or even dominant contribution to spread amplification near the Baltic Sea system at this early stage (Fig.~\ref{fig:PVdiags_36h}b). Spatial patterns are consistent with the patterns we identified for the earlier systems (Sect. \ref{alps-section} and \ref{day2results}). Subsequently, during the second half of day 2, deep convective tendencies persist and divergent and rotational tendencies increase near the system (not shown), suggesting the onset of stage 2 of spread-growth. More importantly, the Baltic Sea system exhibits a clear signature of interaction with the strong PV gradient by its convective outflow, first (at 42h) with positive divergent tendencies occurring predominantly along the gradient near the system. They persist until well beyond 45h (not shown). The Baltic Sea system thereby exhibits a clear signature of the second stage of spread growth according to our interpretation of the three-stage sequence of \cite{Baumgart2019}. In the following, we use the ensemble sensitivity analysis to demonstrate and quantify the relation between variability in the convective outflow and PV variability along the strong PV gradient.

%\subsection{Day 2: ensemble sensitivity and the Baltic Sea convective system}
%\label{sec:balticsystem}
%\subsubsection{Local evolution of associated PV-perturbations} % should get a better name later
\label{sec:BalticPattern}
\begin{figure}
    \centering
    \includegraphics[width=160mm]{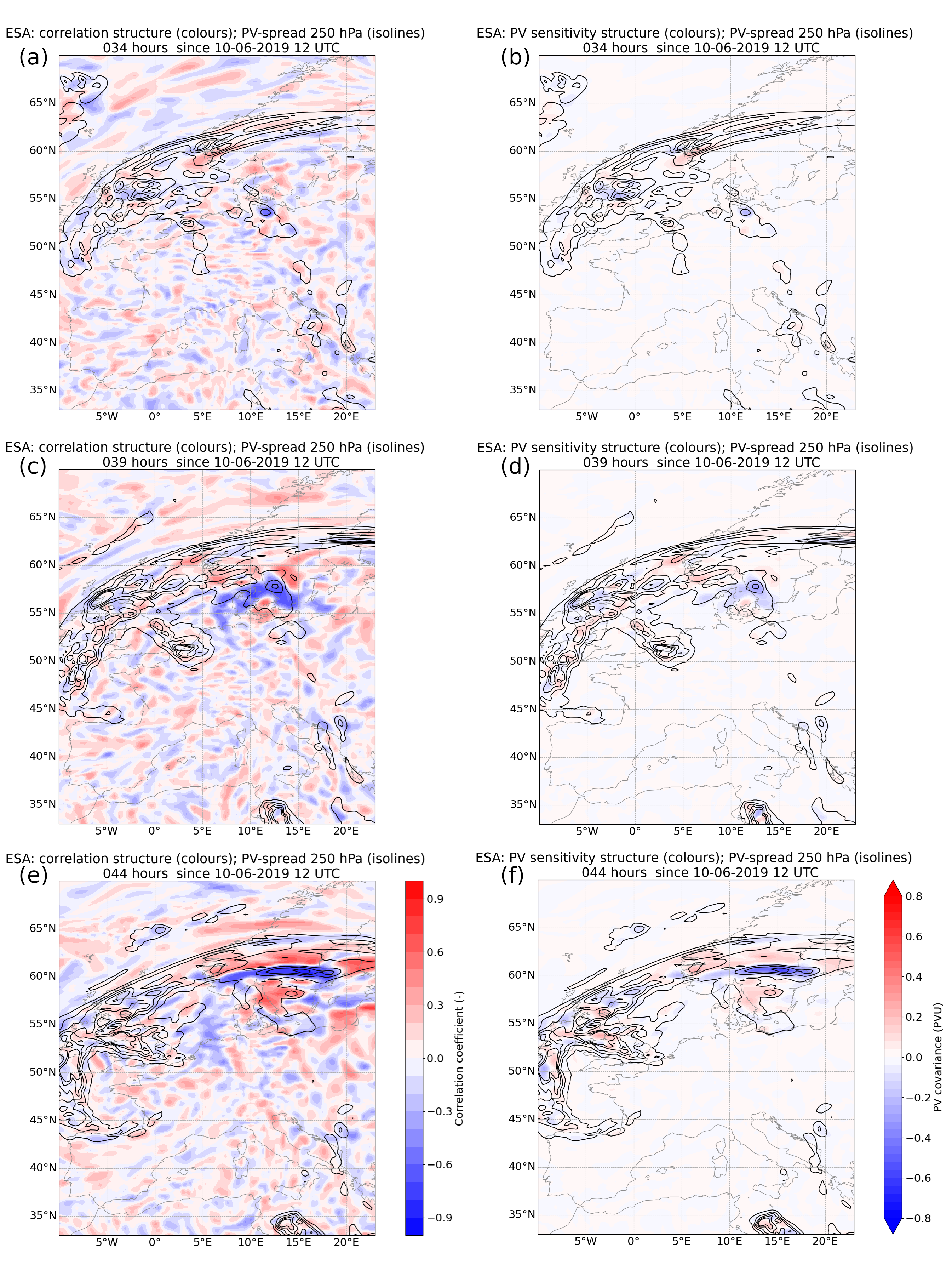}%\includegraphics[width=80mm]{esa-results/ESA-Day2-system-div-t3437_and_t3740/ESA_PV_signal_DAY2_System_t34_to_t37_div_hPa_PV_0068.png}\\
    \caption{Ensemble sensitivity of the potential vorticity at 250 hPa with respect to the mass divergence over the region of the Baltic Sea convective system, 34-37~h. On the left the correlation structure is shown and on the right the PV-amplitude. Top: 34~h, middle: 39~h, bottom: 44~h. Contours show the PV-spread within the ensemble at 0.16 PVU intervals. }
    \label{fig:esa34hours}
\end{figure}
Figure \ref{fig:esa34hours} (b,d,f sequence in the right column) shows the PV deviation associated with enhanced divergent outflow over the convective system that moves from northern Germany into southern Scandinavia. After 34~h, a negative covariance between PV and the divergent outflow strength at 250 hPa arises in a region with about 0.4 PVU spread (Fig. \ref{fig:esa34hours}b); it is co-located with the Baltic Sea system developing over Germany and indicates statistical association of enhanced convective outflow with negative local PV-deviations within the ensemble, which from a PV perspective apparently occurs because of the following sequence. 
%Enhanced generation of negative PV anomalies in the upper troposphere occurs conditional on intenser than average convection and precipitation in a certain ensemble member. 
%This amplified 
When convection and precipitation is amplified in a certain ensemble member, this amplification relates to amplified convective heating in the middle troposphere in a certain simulation (see also Section \ref{sec:precip-systems}). Accordingly, enhanced mid-level heating is within the ensebmle correlated with enhanced upward advection of low PV elsewhere, for instance, towards the jet stream. % gradient are correlated within the ensemble. 
Correspondingly, our growing and amplifying feature with a negative PV deviation indeed moves northward into Southwestern Sweden towards a strong PV gradient (39~h; Fig. \ref{fig:esa34hours}d) and lies along 60 degrees North by 44~h (Fig. \ref{fig:esa34hours}f). The correlations in Fig. \ref{fig:esa34hours} are generally low everywhere at 34~h ($|r|<$0.5), except locally  in proximity of the MCS (Fig. \ref{fig:esa34hours}, left column panels); these areas can be assumed to be affected by outflow of the MCS. At 44~h, ESA correlations $|r|$ are low except for southern Scandinavia.

As an aside we note that, despite the dominant negative deviation PV behaving as expected, positive deviations are also visible in Fig. \ref{fig:esa34hours} after 39 and 44~h. Non-monotonic bands of PV-correlations with the divergent outflow intensity are indeed expected along the gradient where the PV-gradient is not single-signed, travelling in a fixed direction.
The local existence of the double PV-gradients over Central-Sweden and Central-Norway after about 42~h, for instance, is recognised from the folded PV-isolines in Fig. \ref{fig:meanPVpanel}h-i.
The transient interaction with the jet stream's double PV gradient is responsible for PV-dipoles along the jet stream here (not shown).

\begin{figure}
    \centering
    \includegraphics[width=160mm]{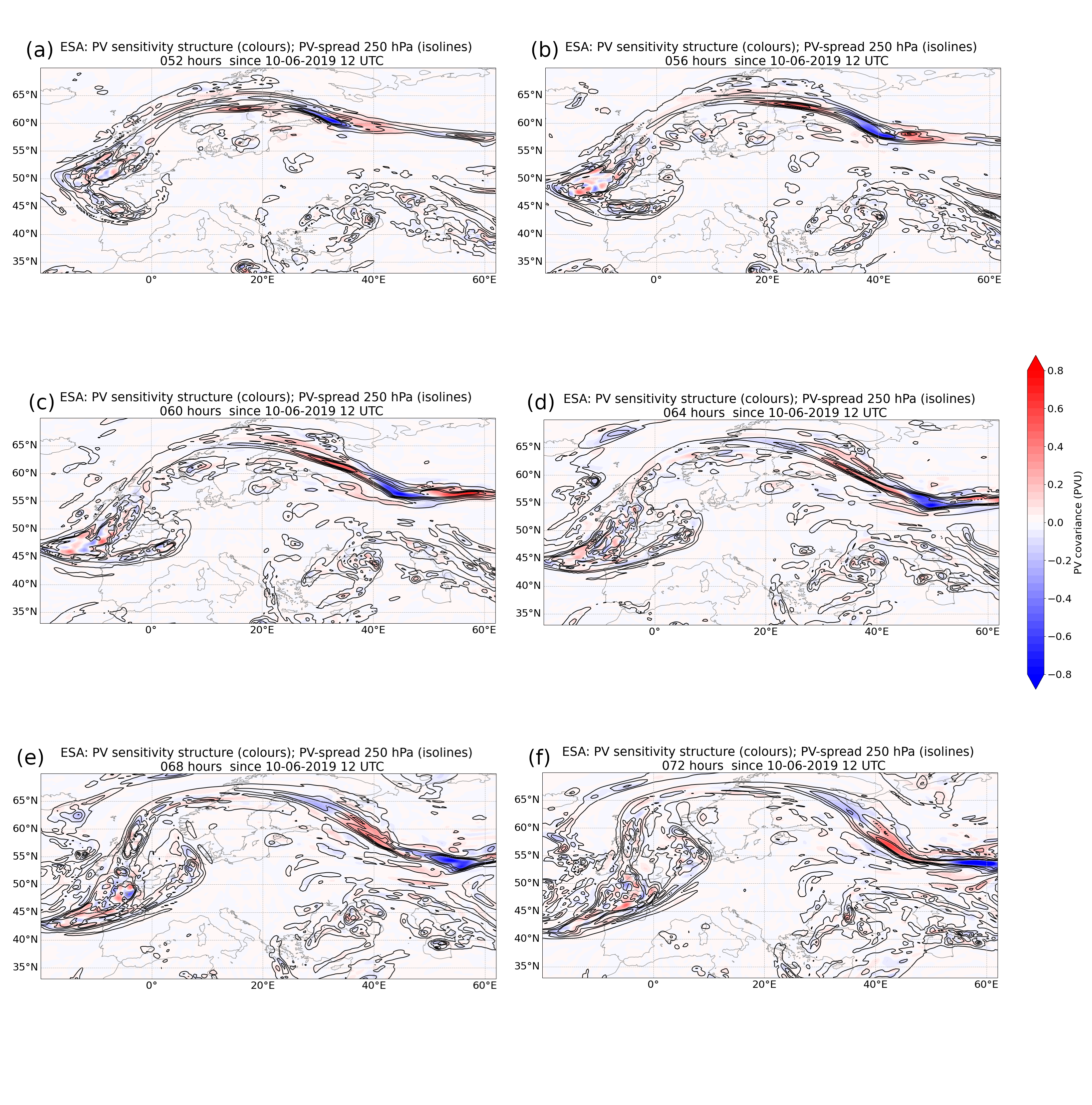}%\includegraphics[width=79mm]{esa-results/ESA-Day2-system-div-t3437_and_t3740/jet-shift/ESA_PV_signal_Day2_System_t34_to_t37_div_hPa_PV_0112.png}\\

    \caption{Potential vorticity co-variance associated with the mass divergence after 34-37~h over the Baltic Sea system's box volume for every 4 hours on the third simulation day. Contours show 250 hPa PV-spread within the ensemble  at 0.16 PVU intervals.}
    \label{fig:jet_shift}
\end{figure}
The PV variability on day 3 as associated with strengthened mass divergence from the Baltic Sea system at 34 to 37~h is displayed in Fig. \ref{fig:jet_shift}. Here, we demonstrate the evolution after the northward jet shift over central Sweden (which occurs at 44~h). While the PV perturbation follows the jet stream downstream for the next 24 hours into Finland and Russia, a wave pattern develops with smaller positive PV-perturbations at both flanks of the negative branch. The positive correlations forming downstream of the negative correlations indicate downstream dispersion akin to a Rossby wave packet. Similar Rossby-wave-like dispersion of spread has been documented by several previous studies \citep[e.g.][]{anwender_etal_2008, riemer_etal_2008, parsons_etal_2019}. Quantitatively, the negative PV deviation associated with the wave pattern has considerable amplitude throughout its residence (within our European nest), with co-variability in PV of up to about 0.8-1 PVU (Fig. \ref{fig:jet_shift}). Furthermore, the local correlation coefficient between upper tropospheric divergence on the previous day and PV at the jet stream's gradient exceeds 0.7 in the core region, linking about half of the PV-variability with the differences in convective (out-)flow rate of the Baltic Sea System. The corresponding association between precipitation rate and the Rossby-wave-like perturbations is investigated in Appendix \ref{prec_var}. In contrast to the Alps System, here, the association is easily tracked downstream with linear analysis. Our correlation pattern maximises within the spread maximum and reaches eastern boundary of our nest by 72h (Fig. \ref{fig:jet_shift}f). 
%Hence, the outflow variability from the Baltic Sea system moves downstream, but in contrast to the Alps System, its outflow variability can easily be tracked downstream with the linear analysis. %The perturbation maximises within the spread maximum and indicates a shifted jet in association with the convective variability, with 
%
In summary, strong correlations overlapping with a shifted jet signify high robustness. Correspondingly, outflow variability projects strongly on the downstream development of the jet stream over Finland and Russia. As hypothesised, from a mechanistic perspective, intenser (divergent) advection following high-amplitude convective heating displaces the jet stream outward (here: northward), followed by Rossby dispersion.

\begin{figure}[t!]
    \centering
    \includegraphics[width=160mm]{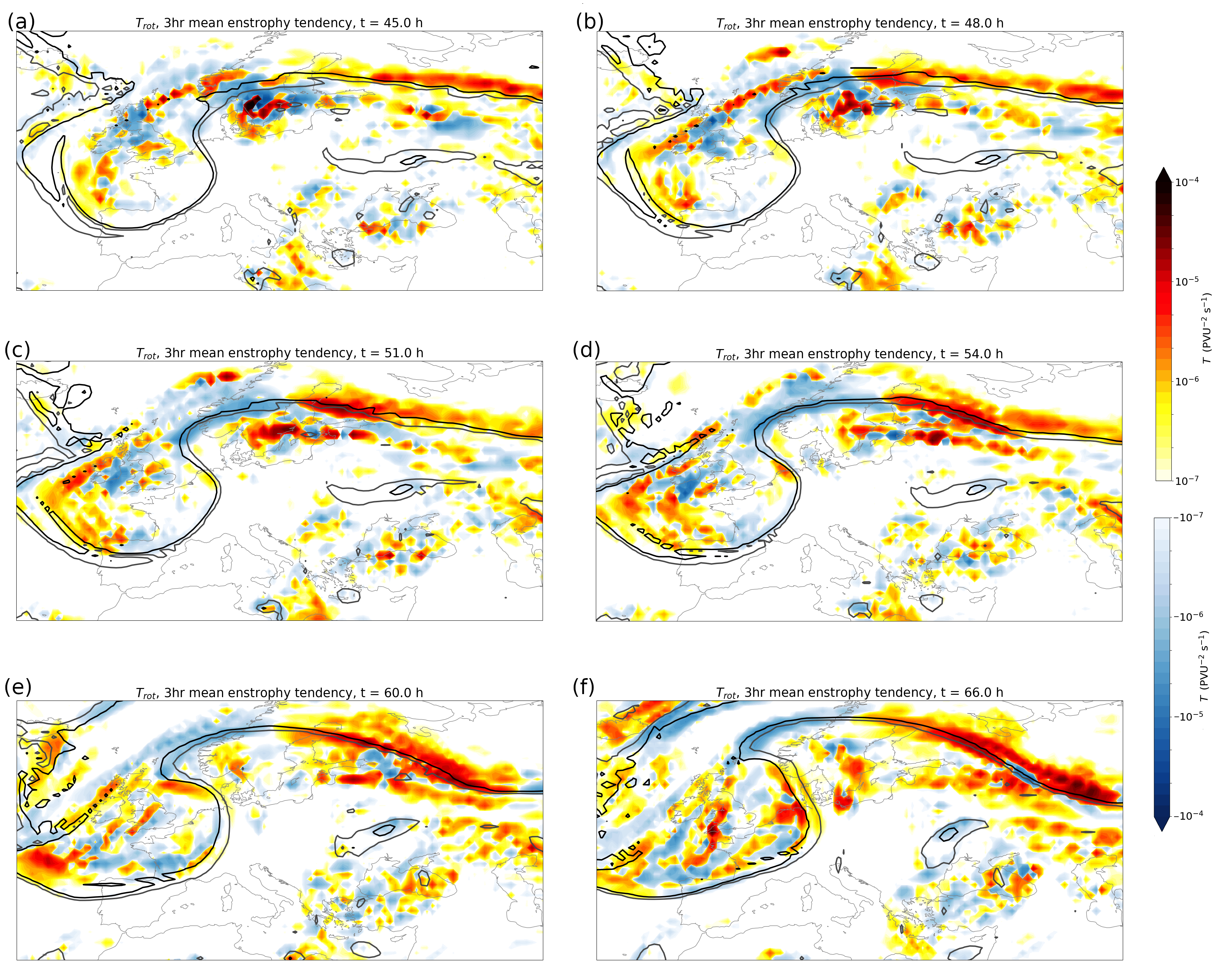}
    \caption{Three hour mean of the potential enstrophy tendency as a result of differential advection by the rotational wind, at $\theta$ of 327.5-335 K: $t=45$, $t=48$, $t=51$, $t=54$, $t=60$ and $t=66$~h. Red colours indicate spread growth, whereas blue colors indicate spread reduction. %Green isolines highlight regions of enhanced PV-spread at 330 K.
    For this level, the dynamical tropopause (2PVU) and 4 PVU contour are also shown in dark grey/black.}
    \label{fig:Trot_day3}
\end{figure}

While propagating downstream, spread further amplifies along the strong PV gradient by nonlinear dynamics. Indicating the onset of a third stage of spread growth, rotational tendencies along the strong PV gradient in the vicinity of the Baltic Sea system first become evident around 45h in the region. Here, rotational tendencies overlap with pre-existing divergent tendencies (not shown) as the convective outflow starts to interact with the PV gradient (Fig.~\ref{fig:Trot_day3}a; cf.~ Fig.~\ref{fig:esa34hours}e,f), which amplifies rotational tendencies strongly in the next 3h and beyond (Fig.~\ref{fig:Trot_day3}b-f). %This local maximum of rotational tendencies subsequently extends into the further downstream region, grows in size, and amplifies along the PV gradient (illustrated between at 51\,h and 66\,h in Fig.~\ref{fig:Trot_day3}c-f). 
The increasing importance of rotational tendencies is in contrast to the evolution of spread downstream of the Germany system. We attribute this distinct difference to the large-scale jet structure. During day 2, the jet exhibited a largely zonal orientation downstream of the Germany system, whereas on day 3 an eminent large-scale wave pattern develops. The larger wave amplitude implies a higher degree of nonlinearity in the underlying dynamics on day 3 - a prerequisite for spread amplification by rotational tendencies \citep{Baumgart2018}. Thereby, \textit{local} spread growth associated with the Baltic Sea system exhibits the characteristics of three-stage spread growth as indicated in the average sense – over large regions and many cases – by \cite{Baumgart2019} and \cite{Selzetal2021}.
\color{black}
%\newline

\subsection{Connection between convective variability and longwave radiation tendencies}
\label{longwave}
%(paragraph literature discussion + open question)
%\citep{TR21} \citep{Baumgart2019} \citep{behrooz}
Near the midlatitude tropopause, PV tendencies associated with longwave radiation are on average the dominant nonconservative PV tendencies \citep{TR21} (in the stages where the advective tendencies dominate).
To a large part, however, these tendencies exhibit relatively small spatial and temporal variation, such that is was proposed that only a small fraction of these tendencies may impact on the synpotic- or smaller scale dynamics \citep{TR21}.  
In the context of an idealized life cycle, \citet{behrooz} demonstrated that cloud-radiative heating exerts systematic differences on the evolution of cyclones, which propagate upscale following the sequence of mechanisms identified by \citet{Baumgart2019}.
In previous spread-growth experiments that considered spread-growth mechanisms averaged over large spatial regions and many cases, however, longwave radiation did not receive particular attention as a major spread-growth mechanism \citep{Baumgart2019,Selzetal2021}. 
%As also mentioned in the previous Subsection, earlier work with the potential vorticity tendency diagnostics has suggested that longwave radiation at near-tropopause levels can be regarded as a relevant process for the amplitude and phase evolution of Rossby Wave Packets \citep{TR21} and idealised cyclones \citep{behrooz}. However,  over the life-cycle of baroclinic wave, it has been regarded as an omni-present and "boring" non-conservative background process \citep{TR21} in trough and ridge amplification (decay) processes - at least when statistically averaged over the extent of a collection of entire troughs/ridges. Furthermore, in another idealised baroclinic life-cycle analysis, by comparing simulations with or without active, all-sky, versus passive, clear-sky, radiative heating, their non-linear dynamics is shown to be relevant for cyclone evolution \citep{behrooz}. \newline
%However, in experiments near the intrinsic limit, the longwave radiation processes have not shown up as domininant contributors to the evolution spread in the statistical average \citep{Baumgart2019,Selzetal2021}. 
Here, our case-specific analysis reveals that longwave radiation may be about as dominant, or locally in space and time even more dominant (at tropopause levels) than direct convective heating tendencies.
This subsection therefore examines, to our knowledge for the first time, spread growth due to longwave radiation in some more depth.
%Therefore, we need to further assess the question how and where these tendencies specifically arise near the intrinsic limit, as well as case-to-case-variability. This section gives an outline to this end. \newline
Before proceeding, it should be noted that the tendencies are expected to magnify whenever anvils with strong (differential) radiation tendencies affect the tropopause, consistently with the analysis of \cite{TR21} and \cite{behrooz}. However, updrafts of not all convective systems penetrate the upper troposphere and tropopause, such that differential longwave heating at these levels only occurs in some convective systems (and with widely varying spatial mean heating rates). Mesoscale convective systems are comparatively likely to produce strong mean longwave heating rates among all convective systems. % , but not in all convective systems as much as in frontal cloud bands: some convective systems do not cause differential heating near the tropopause, but our MCS over Northern Germany on the second day clearly does. %Nevertheless, the divergent winds arising from convection are also not expected to affect the tropopause region as strongly if convection does not reach the upper troposphere (i.e. heating maxima not in the middle troposphere, but lower troposphere). 

\begin{figure}[ht!]
    \centering
    \includegraphics[width=160mm]{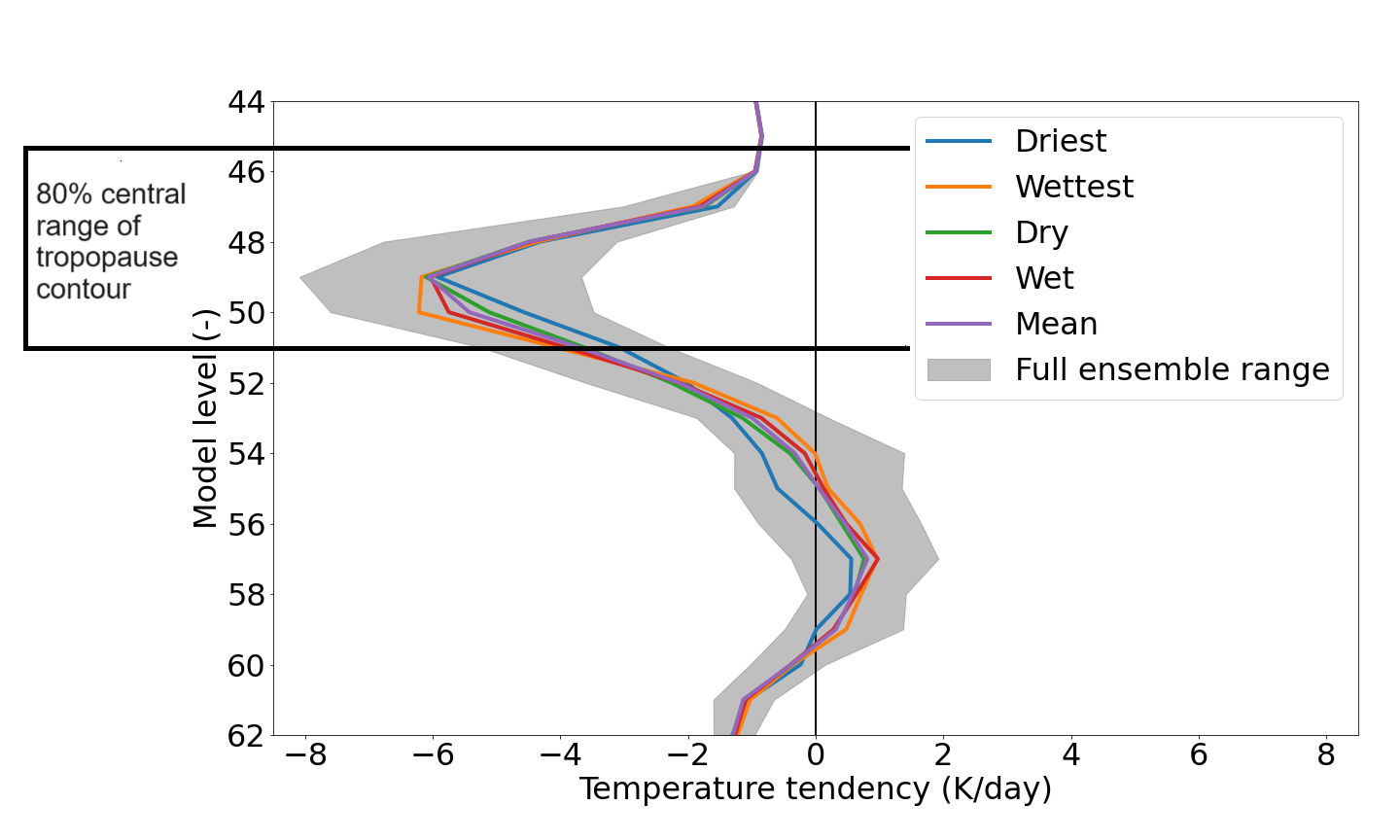}
    \caption{Temperature tendency from longwave radiation as a function of model level over the upper troposphere, after 37.5-38.0\~h in the convective system over Northern Germany. Highlighted are also the model levels at which the dynamical tropopause (2 PVU) is located over the corresponding region. Next to the area and ensemble mean, the area mean over driest (10-14\%), dry (35-39\%), wet (35-39\%) and wettest (10-14\%) ensemble members are shown, which is based on mean precipitation rate over the 35-38\~h time interval. The outer boundary of the grey shading corresponds to ensemble maximum (minimum) of longwave temperature tendencies during the same interval.}
    \label{fig:LW_heating}
\end{figure}

Fig. \ref{fig:LW_heating} shows the temperature tendencies originating from the longwave radiation parameterisation over the region of the convective system in Northern Germany after 38\~h. It shows that there is some connection between the temperature tendencies at anvil level (model levels 48-51; near-tropopause, 200 hPa) and the precipitation rate (35-38\~h). The intenser the convective precipitation is, the larger the mean temperature tendencies tend to get at the anvil - especially at level 50. When we look further down at levels 55-59 (between 250 and 300 hPa) the pattern is reversed.

It is important to realise that the effect of these heating tendencies onto potential vorticity spread growth mostly derives from variability in \textbf{vertical gradients} of potential temperature tendencies. The slightly amplified vertical gradient in Fig. \ref{fig:LW_heating} hence suggests amplified negative PV-perturbations stemming from longwave tendencies just below anvil level, at model levels 51-53 in all members and by extension levels 49-54 in some members (an overlap with the dynamical tropopause). On the contrary, in all members, (weak) positive perturbations will occur near level 60 and (stronger) near level 47. The rough scaling of the heating gradient at levels 50-53 with precipitation rate on the one hand, and subtle vertical shifts of cooling/heating on the other hand, could together likely explain the strong tendencies originating from longwave radiation in the tendency diagnostics - at least to some extent.

Interestingly, a comparison of the tendencies between 35-36 and 38\~h suggests that a more elevated overshoot in the Baltic Sea convective system at model levels 48-49 tends to be associated with the wet and wettest ensemble members. Furthermore, this relation between the area mean vertical longwave radiative heating gradient at these levels and the precipitation rate shows that the convective variability (both precipitation rates and outflow rates) are linked with longwave radiation tendencies. Therefore, convective variability has the potential to not only trigger PV-spread tendencies induced by the convective parameterisation, but also by differences in longwave radiation. 

Although our experiment has not been designed to target extensive analysis of longwave radiation, our analysis reveals an interesting and close link between longwave radiative heating variability and variability in the mesoscale convective system on day 2 of our experiments, which has been shown to affect spread growth. Therefore, further investigations of case to case variability in longwave and convective heating tendencies near the intrinsic limit would be beneficial to understand the link of longwave spread tendencies to convective variability and spread-growth stages better, ideally with refined resolution in the upper troposphere and lower stratosphere.

%\color{red}We could discuss whether the LW term is more of a redistributor term w.r.t. divergent outflows (rather than an amplifier/probably more so: weak destructor), but this goes well beyond our scope and could hence be ignored.\color{black}
%\color{red}Don't forget to add the relevant literature discussion as well!\color{black}
%
%
%
\section{Summary \& discussion}
We investigate spread growth in an ICON ensemble experiment initialised with miniscule initial-condition uncertainty (rescaled to 0.1\,\% of a typical current-day operational level).
The experiment covers a three-day period in summer, when three mesoscale convective systems develop over Europe in the vicinity of the midlatitude jet.
Even in this setup near the intrinsic limit of predictability, one of the convective systems substantially affects the synoptic-scale flow about one day after its convective initiation.
Variability in the jet location of 
$\approx 100\,$km is associated with variability of upper-tropospheric divergence and, correspondingly, 3-4\,\% uncertainty in precipitation rates of this system. Precipitation variability in an operational ensemble is currently larger by about an order of magnitude. Consequently, for (almost) any realistic forecast, convective uncertainty amplifies spread growth very rapidly, whereby stage 3 of spread growth (typically) dominates immediately in an ensemble \citep[see also][]{Selzetal2021}. %Hence, precipitation variability within an ensemble of real forecast models is much (an order of magnitude) larger than here. 
Therefore, even without regime transition in the dominant spread growth dynamics from stage 1 to 3, enhanced precipitation variability within realistic ensembles
would likely permit much larger
jet stream uncertainty at short lead times.

The focus of the study is on the analysis of the conceptual three-stage sequence of dominant spread-growth mechanisms, previously identified from a statistically averaged perspective by \cite{Baumgart2019, Selzetal2021}, i.e., as composite evolution over many cases and very large regions.
More specifically, the key question of this study is: Can we identify the same three-stage sequence \emph{locally}, i.e., in the spread evolution originating from specific convective systems? 
\subsection{Stages of spread growth}
The spread evolution originating from the individual convective systems exhibits some commonalities, but also distinct differences.
The commonalities can be summarized as follows. 
Convective PV-spread tendencies dominate spread growth in the initial stages of the convective systems, which can be expected due to the design of the experiment.
Longwave-radiative spread tendencies become equally or even more important in the convective region (and partly downstream) a few hours later. 
The upper-tropospheric divergent flow amplifies spread in the vicinity of the convective systems, and all systems interact with the strong PV gradient associated with the jet.

Only one of the three convective systems (the Baltic Sea system) exhibits the full signature of the previously identified three-stage sequence.
After initial growth of spread, which is dominated by convection, variability in the divergent outflow from the Baltic Sea system interacts with the jet and dominates amplification of PV spread along the strong PV gradient in the interaction region.
Subsequently, the spread pattern propagates downstream and can unambiguously be traced by (linear) ensemble sensitivity analysis.
Spread amplification in the downstream region is dominated by nonlinear growth, as indicated by rotational tendencies aligned with the strong PV gradient.

For another convective system (the Germany system), divergent tendencies also dominate after an initial convective stage.
In contrast to the Baltic Sea system, however, the associated spread amplification occurs locally above the convective system, arguably associated with the adjustment of convective outflow to a balanced, anticyclonic flow component, rather than in direct interaction with the jet.
Subsequently, an interaction of this anticyclone with the jet is indicated by rotational tendencies along the strong PV gradient. 
This evolution is thereby consistent with the second stage of the conceptual model proposed by \cite{Zhang2007}, which assumes that a general adjustment-to-balance process occurs after a first stage of convective spread growth (and before a third stage of growth on geostrophically-balanced scales). In our case, however, this interaction with the jet stream is not followed by substantial jet variability in the associated region downstream.

For the third system, the Alps system, both PV-spread tendencies and ensemble sensitivity analysis indicate an interaction of convective outflow with the jet.
Subsequent nonlinear growth is indicated by rotational tendencies.
However, neither of these two processes dominates spread growth for any period of time.
Neither distinct stages of spread growth can thus be identified after the initial convective growth, nor a persistent coherent downstream sensitivity with respect to the system's outflow.

In addition to the spread-growth mechanisms emphasized in the previous multi-stage conceptual models by \cite{Zhang2007} and \cite{Baumgart2019},
we find that longwave-radiative tendencies make a considerable contribution to spread amplification in the vicinity of convective systems.
Briefly after the start of the first convective stage of spread growth, longwave-radiative tendencies become comparable or even larger than convective tendencies.
Importantly, longwave-radiative tendencies can outlive convective tendencies and extend over larger regions, arguably associated with the anvils of the convective systems.
While a considerable contribution by longwave-radiative tendencies has been documented as an aside previously \citep[e.g.,][]{Baumgart2019, Selzetal2021},
we here demonstrate explicitly that variability of longwave-radiative tendencies in one of our convective systems is associated with local convective variability.
In this sense, longwave-radiative tendencies communicate convective variability to the tropopause region and can be thereby considered as an inherent contribution to the convective stage of spread growth. 
Moreover, by extending over larger regions, longwave-radiative tendencies arguably contribute to an upscale impact of convective variability, complementing the role of the convective tendencies in the first stage and divergent wind in a second conceptual stage of spread growth.

On the one hand, substantial divergent tendencies appear very early (before 12~h) \citep[consistently with][]{Zhang2007,Baumgart2019}, indicating an onset of stage 2 at the PV gradient near the Alps System. On the other hand, however, another region becomes saturated with convective and longwave tendencies at 36~h, which follows on an episode with little spread, indicating that stage 1 is locally re-initiated much later.
For both systems, the initiation of stage 1 is rapidly followed by the onset of stage 2 and by rotational spread tendencies (conceptually indicating the onset of stage 3). The rotational tendencies of the conceptually exemplary Baltic Sea system appear after about 12~hour of stage 1 and, subsequently, stage 2 spread growth (at about 48~h, Fig.~\ref{fig:Trot_day3}b).
The variation in the timing of the onset of stage 1, followed by a rapid transition to subsequent stages, contrasts estimates from the mean perspective and certainly indicates substantial variation in spread-growth stages locally. From the mean perspective variability cannot be determined easily.
The differences may be explained by temporal variability in the development of rapidly spread-saturated areas at larger mesoscales; convective initiation occurs at random times an, subsequently, convective-scale spread can saturate in a couple of hours \citep[e.g.][]{hohenegger2007,Weyn_Durran_2017,Groot_Tost_2022} in regions of active convection. %While convection initiates at random times, mesoscale spread-saturation occurs at random times, w
Thereby, spread-prone regions may locally project spread on jet streams quickly and feed larger scale spread (like the Baltic Sea system). 
Finally, few of the MCSs in proximity of the jet stream likely associate with lagged downstream jet variability (and hence, traces of their variability likely decay), i.e., spread evolves rarely across all three conceptual stages (whether it does may incidentally be probabilistic \citep[e.g.][]{Lorenz1969,melhauser_zhang_2012}). 

%Bearing in mind that we find considerable variability in the onset times of any conceptual spread-growth stage, we have to consider any spread evolution an exemplary realisation of a probability distribution of spread evolution. In this consideration, spread initially grows from the convective stage, with possibility to reach the later two conceptual stages. Nevertheless, few of MCSs in proximity of the jet stream likely associate with lagged downstream jet variability (and hence, traces of their variability likely decay), i.e., spread evolves rarely across all three conceptual stages.\\

%Therefore, a 
A key question that our results imply is:
What governs the observed variability in the characteristics of the presumed second and third stages?
Based on a case study of merely three convective systems, we can provide a preliminary answer at best.
We attribute the lack of spread growth during the presumed occurrence of third-stage spread-growth dynamics associated with the Germany system to the jet structure in the downstream region.
The associated strong PV gradient approximately exhibits a zonal orientation, which can be expected to support a linear evolution of (small) perturbations to the jet.
Nonlinear dynamics, however, are required for spread to amplify due to rotational tendencies along a jet \citep{Baumgart2018}.
Consistently, third stage spread growth due to rotational tendencies is observed downstream of the Baltic Sea system in a higher-amplitude wave pattern.
For the Alps system, the upstream trough is undergoing rapid wave breaking. 
Arguably, this wave breaking prohibits the coherent downstream propagation and amplification of spread along the jet.
Moreover, spread tendencies that are not related to the  Alps system (associated with a weather system over the UK and the North Sea; not discussed in detail in the results section) arguably complicate the attribution of the spread evolution within the trough to the Alps system.

An explanation for the observed variability in the presumed second stage of spread growth is more elusive.
To our knowledge, the linear, time-dependent analytical model for geostrophic adjustment and its application to idealized experiments of upscale growth by \cite{shutts1994,Bierdel2017, Bierdel2018} provides the most comprehensive theoretical framework for upscale growth during a presumed second stage.
While the model generally confirms the hypothesis by \cite{Zhang2007} that geostrophic adjustment is a viable mechanism for upscale growth, the model does not consider the existence of strong, localized PV gradients and thereby it is not clear how the model could be applied to investigate the flow dependence of geostrophic adjustment in the vicinity of a jet.
%While a linear, analytical model for geostrophic adjustment applied to idealized experiments \citep{Bierdel2017, Bierdel2018} confirms the hypothesis by \cite{Zhang2007} that geostrophic adjustment is a viable mechanism for upscale growth, this model does not consider the existence of strong, localized PV gradients and thus it is not clear how the model can be applied to investigate the flow dependence of geostrophic adjustment in the vicinity of a jet.
%While the model generally confirms the hypothesis by \cite{Zhang2007} that geostrophic adjustment is a viable mechanism for upscale growth, the model does not consider the existence of strong, localized PV gradients
In our experiment, the differences between the Germany system and the Baltic Sea system may relate to differences in the distance of the convective system to the jet (arguably relative to the local deformation radius, which is an upper bound to the scale of geostrophic adjustment).
Regarding the Alps system, we may speculate that the lack of a distinct second stage relates to the alignment of the interaction between the convective variability and the jet within the wave phase of the midlatitude wave.
The Alps system interacts with the upstream trough, whereas spread growth due to divergent outflow from the other two systems occurs within the downstream ridge.
The downstream impact of recurving tropical cyclones has been shown to depend sensitively on the relative position of recurvature with respect to the midlatitude wave pattern \citep{keller2019}.
%To the extent that this result transfers to mesoscale convective systems, albeit it needs to be noted that recurving tropical cyclones exhibit considerably stronger divergent outflow, we may hypothesize that the spread growth due to convective systems may equally depend on their relative location to the midlaitude wave pattern.
To the extent that this result may transfer to mesoscale convective systems, we could hypothesize that spread growth due to convective uncertainty depends on the ''phase relation'' between convective systems and the midlatitude wave pattern, %(and correspondingly, timing of the occurrence), 
and not only on distance.
%Lastly, the Alps system during its life time may partially intersect with the PV gradient, which presumably could also complicate the development of the second stage of spread growth. 
Clearly, further research into the dynamical constraints under which the second and third stage is favoured (or, conversely, restricted) is needed to verify our rather speculative explanations.

%Both the Germany and the Baltic Sea system move towards the jet within the ridge over central Europe.
%The observed differences in stage 2 of the two systems may thus simply be related to the distance from the jet: Based on the figures that we present in Sect.~\ref{sec:results}, we roughly estimate that the Baltic Sea system approaches the strong PV gradient association with the jet to within 300\,km, whereas the Germany system only to within 500\,km.

%
%
\subsection{Reflection on the experimental design and outlook}
We find that spread evolution may be interpreted more robustly under some conditions when spread tendency diagnostics are complemented with a linear sensitivity analysis. By combining the tools, the quantitative impact of a precursor on spread downstream is established. Here, about half the local downstream jetstream variability at 60~h relates to the Baltic Sea system's outflow. While linear analysis does not allow to robustly quantify statistical connections with substantial nonlinearity, we utilise the improved tractability of convective variability by combining careful experimental design with knowledge of the close association between outflow and convective heating \citep{Grootetal2023}.

In a convection-permitting configuration the downstream impact of outflows of organised convection %, and their effect on flow uncertainty, 
can be traced in more detail, because %. Since 
the representation of these outflows is refined in high-resolution simulations \citep{Grootetal2023,Groot_Tost_2022b, grootthesis}. In such simulations, the outflow can be understood conceptually as a combination of two components: i)
%\begin{itemize}
%    \item 
a contribution linearly dependent on latent heating 
%    \item 
and ii) an approximately multiplicative component resulting from convective organisation or geometry. Nonetheless, from a practical point of view, this will first require a development to track each of the above two components of outflow variability, which would require a robust quantitative tracking framework (this needs further investigation, following \cite{Grootetal2023}). % have shown that the second components cannot be isolated easily in ICON their convection-permitting configuration, which means that unambiguous parameters to quantify the second component during a number of cases with variation in convective organisation are needed first. 
%\end{itemize}%and an effect of convective, 
Non-linear (multiplicative) effects reduce the tractability of convective variability downstream when both components are represented, i.e., at high resolution. Furthermore, both effects are time-dependent in real systems. Consequently, it is likely difficult to track non-linear variability associated with outflow variability in convection-permitting configurations. %Interactions within the gravity wave spectrum both complicate analysis and improve representation \citep{Grootetal2023,grootthesis}. 

\cite{Grootetal2023} have shown that grid spacing affects the representation of divergent outflows, differing between convection-permitting 1 km grid and convection-parameterising 13 km grid configurations of ICON. 
Here, %we have utilised the expected 
the near-linear response of divergent outflows in parameterised convection allows us to effectively track the role of the convective variability (and tendencies) downstream. Accordingly, by investigating a configuration with near-linear response, convective variability and divergence are tightly related. Future experiments could aim to identify propagation of individual wave signals, as associated with convective outflows, and attribute the waves to divergent winds. Thereby, a wavelet analysis of the winds, and possibly even spread tendencies, could be utilised to further establish their connection, although this may be very challenging from a practical perspective. Furthermore, it will be worthwhile to examine the co-variability of convective activity and longwave radiative tendencies
%the connection between longwave radiative tendencies and divergent outflows could be explored 
in further quantitative and qualitative detail in future work.
% between the two. Subsequently, we could speculate that interference between convectively originating wave signals, while affecting mass divergence non-linearly at given latent heating rates, may be identified. 
%Similarly, idealised connections between the representation of convective variability and the representation of longwave radiation in models might also be worth investigating, as well as its potential to quantitatively impact convective outflow rates.  
%A lot of insight can be gained from the simplified configuration. 
%
%To us, potentially the most fruitful avenue for future research is to better understand the case-to-variability of spread growth during the presumed second and third stages.
%In particular investigating the second stage, which implies the upscale impact of convection, can be expected to benefit from using convection-permitting experiments.  
%
%
Finally, we think that further understanding of case-to-case variability of spread growth during the presumed second and third stages is a particularly fruitful avenue for future research, especially if we can constrain dynamical conditions favouring strong spread growth.
%In particular 
Investigating the second stage, which implies the upscale impact of convection, can be expected to specifically benefit from using convection-permitting experiments.  
\appendix
\section{Relation jet stream spread and precipitation rate spread}
\label{prec_var}
We can assume that correlation between upper troposhperic divergence variability and (downstream) PV-perturbations is caused by differential divergence, which results from differences in column latent heating. If this is true, the correlation signals found in Section \ref{sec:BalticPattern} can also be diagnosed in an ensemble sensitivity analysis with respect to precipitation rates. This is strongly suggested by the findings of \cite{Baumgart2019,Groot_Tost_2022b} and \cite{Grootetal2023}.\newline
An ensemble sensitivity analysis of the instantaneous precipitation rate over the Baltic Sea system complements the sensitivity analysis of the mass divergence rate, because upper tropospheric divergence rate is expected to relate linearly to precipitation rate variability across the ensemble (for our case and configuration; see \cite{Grootetal2023}). Therefore, we can test implicitly whether box-averaged latent heating perturbations trigger differential divergent outflow winds, which later cause differential advection of PV perturbations further downstream and modify the PV-field.% and producing forecast uncertainty near intrinsic limit \citep{Baumgart2019}. 

\begin{figure}
    \centering
    \includegraphics[width=150mm]{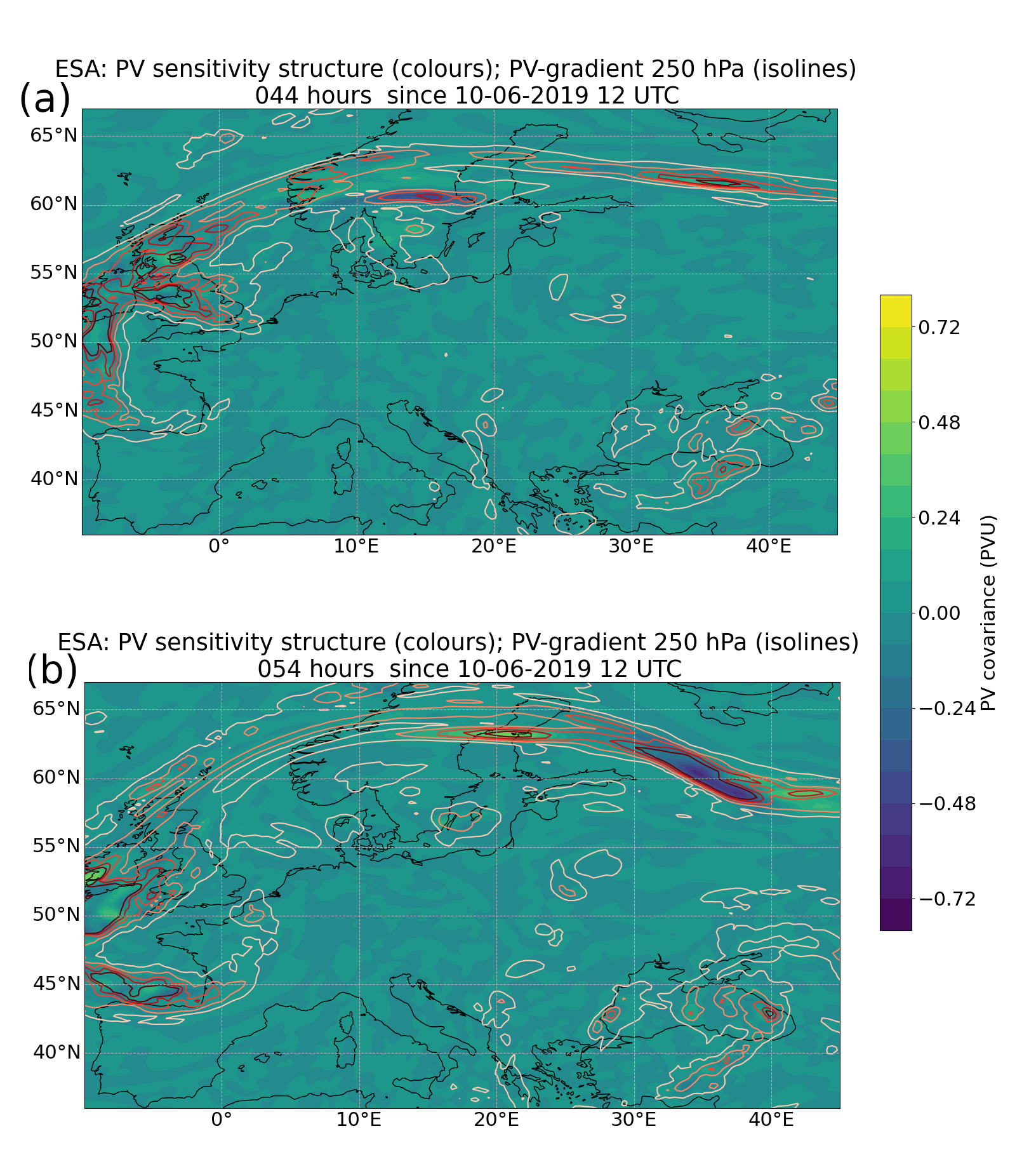}\\
    \caption{Ensemble sensitivity of the potential vorticity at 250 hPa with respect to the precipitation rate of the Baltic Sea convective system at 35-38~h. Top: +44~h%, middle: +49 hours,
    ; bottom: +54~h. Isolines indicate the magnitude of PV-spread at 250 hPa at 0.16 PVU intervals.}
    \label{fig:precipESA}
\end{figure}

Fig. \ref{fig:precipESA} shows the ensemble sensitivity of 250 hPa potential vorticity to precipitation rate variability of the Baltic Sea system at +35 to +38h. This is the time interval displaying the strongest downstream correlation about 15-20 hours later, near the PV gradient. Furthermore, the system strongly develops after 35h. An elongated feature with of a negative PV-perturbation just north of 60 $^{\circ}$ N is visible in Fig. \ref{fig:precipESA} (see also Fig. \ref{fig:esa34hours}f). %, after 44 hours (see also Fig. \ref{fig:esa34hours}). 
This feature grows into a larger-scale wave-like perturbation at %+49 hours and forms the wave pattern over Russia at 
54~h, with elongated bands of positive PV-perturbations with zonal orientation to the west and east of the dominant PV-perturbation, which is negative. \newline
The signal confirms the hypothetical relation between variation in the latent heating and downstream PV-perturbations at 250 hPa. % in the upper troposphere 
Divergent outflows are a key mediator during a cascade of underlying spread-growth processes (Sect. \ref{36hoursdiscussion}). %The relation seems to be imposed by the diabatic generation of negative PV-perturbations associated with the convective system, enhancing divergent outflows containing negative PV-perturbations. The perturbed air mass associated with these outflows is subsequently transported towards the PV-gradient and induces differential advection across the ensemble there. It is suggested by the wave patterns, found here and in Section \ref{36hoursdiscussion}, that the Lorenz type of spread growth starts to impact the evolution of the ensemble difference winds: difference winds cause differential advection of PV that is purely dynamical. \newline
%In the following  Section, the perspective from the PV tendency diagnostics will be discussed. \newline
%The precipitation signals after 35-38 hours and 42-46 hours turn out to be correlated with about $r=-0.5$. This negative correlation explains why the ensemble sensitivity in terms of 250 hPa PV is anti-correlated between 34-37 and 43-46 hours: ensemble members with very intensive convection initially tend to have less intensive convection during the later of the two time intervals, on average. 
Given the linear correlation ($R^{2}$) of about 0.8 between mass divergence variability of the Baltic Sea system and its precipitation rate variability within the ensemble (not shown), the signal of Figs. \ref{fig:esa34hours}f and \ref{fig:jet_shift}) is only slightly diluted in Fig. \ref{fig:precipESA}. % when looking at the ESA of the precipitation rate. %\color{red}Assuming causal interference \color{blue}refer to Mirjam's paper and Kretschmer et al or ref therein if needed\color{red}and approximately multiplying the signal of mass divergence in the ensemble sensitivity analysis with the precipitation-mass divergence correlation, the signal would resemble Figure \ref{fig:precipESA} to some extent. This is a speculative computation. On the other hand, strong opposing signals in Figure \ref{fig:43to46hours} in combination with the weak anti-correlation between precipitation in the two time intervals would suggest that here a common driver more likely explains the covariance. Furthermore, 
In short, our statistical analysis of the Baltic Sea system demonstrates that mass divergence is occasionally an important mediator in the evolution of flow spread, which linearly projects latent heating variability (through differential advection) onto the flow further downstream. %==\> All a little speculative, a few more checks needed, maybe move to the discussion and explain it better. \color{black}
\newpage
\section*{Supplementary manuscript metadata and info}
\subsection*{Acknowledgements}
The authors would like to thank Tobias Selz, Holger Tost and George Craig for their feedback on the initial idea for this study and for the joint discussions within the A1 project. Furthermore, we would like to thank Tobias Selz for providing further technical support, e.g., by providing case-specific initial conditions following \citet{Selzetal2021}, providing access to the associated code and suggestions to improve the work.
\subsection*{Funding}
The research leading to these results has been done within the subproject ‘A1 - Multiscale analysis of the evolution of forecast uncertainty’  of the Transregional Collaborative Research Center SFB / TRR 165 ‘Waves to Weather’ funded by the German Research Foundation (DFG). EG acknowledges additional funding at the University of Oxford by the Leverhulme Trust grant funding EG's current position.  \newline
The authors would also like to acknowledge the computing time granted on the supercomputer MOGON 2 at Johannes Gutenberg-University Mainz (hpc.uni-mainz.de, last accessed: 12-02-2025) and Wavestoweather's computational infrastucture.
\subsection*{Conflict of interest statement}
The authors declare that they are not aware of any conflict of interest. 
\subsection*{Author contributions}
This study has been initiated by EG. EG carried out the ICON simulations, did the initial analysis and interpretation, made figures and started working on an initial draft. MR has assisted with writing and has been deeply involved in the follow-up discussion, leading to substantial progress and revisions. The overall work is a joint interpretation of this study. 
\subsection*{Data availability statement}
The most important data associated with this work are available via \textit{https://doi.org/10.5281/zenodo.14861019}, last accessed: 12-02-2025.
%\begin{itemize}
%    \item Show evolution ESA of %precipitation variability
%    \item Explain possibly anti-correlation first and second half of the convective system's evolution; correlation coefficients second system, 35 to 38h versus 42 to 46h are -0.5
%    \item (result suggests common driver, but obviously not causal!! by multiplication of divergence correlations)
%\end{itemize}
\newpage
 \bibliographystyle{plainnat}
 \bibliography{bibl.bib}
\end{document}